\documentclass[10pt,conference]{IEEEtran}
\usepackage{cite}
\usepackage{amsmath,amssymb,amsfonts}
\usepackage[indLines=true, noEnd=false]{algpseudocodex}
\usepackage{algorithm}
\usepackage{graphicx}
\usepackage{textcomp}
\usepackage[hyphens]{url}
\usepackage{fancyhdr}
\usepackage{hyperref}
\usepackage{physics}
\usepackage[shortlabels]{enumitem}
\usepackage{mdframed}
\usepackage{pifont}
\usepackage{multirow}
\usepackage{ulem}

\providecommand{\sbf}[1]{\textsf{\textbf{#1}}}

\providecommand{\revise}[1]{\textcolor{black}{#1}}

\definecolor{mplblue}{HTML}{1f77b4}
\definecolor{mplred}{HTML}{d62728}

\newcommand{\hpcayear}{2025}

\newcommand{\name}{HATT}

\newcommand{\had}{\hat{a}^\dagger}
\newcommand{\ha}{\hat{a}}

\newcommand{\hpcasubmissionnumber}{297}
\title{HATT: Hamiltonian Adaptive Ternary Tree for Optimizing Fermion-to-Qubit Mapping}

\def\hpcacameraready{} 

\newcommand\hpcaauthors{Yuhao Liu$^\dagger$, Kevin Yao$^\dagger$, Jonathan Hong$^\dagger$, Julien Froustey$^\ddagger$, Ermal Rrapaj$^\parallel$, Costin Iancu$^\parallel$, Gushu Li$^\dagger$, Yunong Shi$^\S$}
\newcommand\hpcaaffiliation{University of Pennsylvania$^\dagger$, University of California, Berkeley$^\ddagger$,  Lawrance Berkeley National Lab$^\parallel$, AWS Quantum Technologies$^\S$}
\newcommand\hpcaemail{liuyuhao@seas.upenn.edu, keyao@seas.upenn.edu, johon@seas.upenn.edu, jfroustey@berkeley.edu, \\ermalrrapaj@lbl.gov, cciancu@lbl.gov, gushuli@seas.upenn.edu, shiyunon@amazon.com}

\author{
  \ifdefined\hpcacameraready
    \IEEEauthorblockN{\hpcaauthors{}}
      \IEEEauthorblockA{
        \hpcaaffiliation{} \\
        \hpcaemail{}
      }
  \else
    \IEEEauthorblockN{\normalsize{HPCA \hpcayear{} Submission
      \textbf{\#\hpcasubmissionnumber{}}} \\
      \IEEEauthorblockA{
        Confidential Draft \\
        Do NOT Distribute!!
      }
    }
  \fi 
}

\fancypagestyle{camerareadyfirstpage}{%
  \fancyhead{}
  
  \fancyhead[C]{}
  \fancyfoot[C]{}
}
\begin{document}





\maketitle

\ifdefined\hpcacameraready 
  \thispagestyle{camerareadyfirstpage}
  \pagestyle{empty}
\else
  \thispagestyle{plain}
  \pagestyle{plain}
\fi

\newcommand{\hpcaheight}{0mm}
\ifdefined\eaopen
\renewcommand{\hpcaheight}{12mm}
\fi



\begin{abstract}

This paper introduces the Hamiltonian-Adaptive Ternary Tree (\name{}) framework to compile optimized Fermion-to-qubit mapping for specific Fermionic Hamiltonians. In the simulation of Fermionic quantum systems, efficient Fermion-to-qubit mapping plays a critical role in transforming the Fermionic system into a qubit system. \name{} utilizes ternary tree mapping and a bottom-up construction procedure to generate Hamiltonian aware Fermion-to-qubit mapping to reduce the Pauli weight of the qubit Hamiltonian, resulting in lower quantum simulation circuit overhead. Additionally, our optimizations retain the important vacuum state preservation property in our Fermion-to-qubit mapping and reduce the complexity of our algorithm from $O(N^4)$ to $O(N^3)$. Evaluations and simulations of various Fermionic systems demonstrate \revise{$5\sim20\%$} reduction in Pauli weight, gate count, and circuit depth, alongside excellent scalability to larger systems. Experiments on the Ionq quantum computer also show the advantages of our approach in noise resistance in quantum simulations.
 
\end{abstract}

\section{Introduction}\label{sec:introduction}

Simulating Fermionic systems, composed of particles such as electrons, protons, and neutrons~\cite{particle2022review}, is an essential application area for quantum computing. These systems are integral to various critical physics models, including the electronic structure of molecules in quantum chemistry~\cite{mcquarrie2008quantum}, the Fermi-Hubbard lattice model in condensed matter~\cite{altland2006condensed}, neutrino evolution from astroparticle physics~\cite{barger2012physics,Patwardhan:2020,Cirigliano:2024}, etc. Being able to simulate large scale Fermionic systems could directly benefits material science, drug discovery, etc. However, Fermionic systems are intrinsically quantum and hard to simulate on classical computers on a large scale due to their exponential and super-exponential complexities. It can take tens of millions of cores in a supercomputer to perform ab initio electronic structure simulation~\cite{hu2022performance}, and it is getting harder to scale larger, limiting the scales scientists could study. Thus, using quantum computers to simulate Fermionic systems provides a crucial solution to the complexity barrier as quantum computers can simulate these systems with linear resource requirements~\cite{feynman1982simulating}.



\begin{figure}[t]
    \centering
    \includegraphics[width=\linewidth]{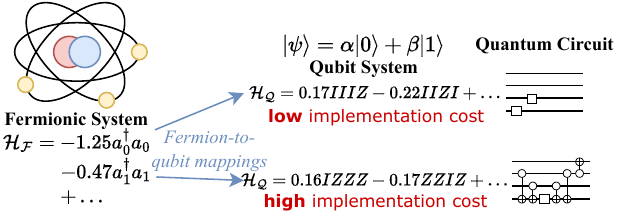}
    \caption{Overview of simulating Fermionic systems with quantum computers. The mapping on the upper half is better than that in the lower half because it generated a qubit Hamiltonian with fewer Pauli operators, and the circuit implementation cost is lower.}
    \label{fig:simulation-overview}
\end{figure}

To simulate a Fermionic system on a quantum computer requires an essential step, the Fermion-to-qubit mapping.
This is because most quantum computers are built with qubits with different statistical properties than Fermions.
A Fermion-to-qubit mapping bridges this gap by creating the Fermionic statistics in a qubit system.
As depicted in Figure~\ref{fig:simulation-overview}, a Fermionic Hamiltonian $\mathcal{H}_\mathcal{F}$ (on the left) is usually expressed by the creation and annihilation operators $\{a_i^\dagger, a_i\}$, which do not naturally arise in a qubit system.
A Fermion-to-qubit mapping will find Pauli strings to represent the creation and annihilation operators, converting the Fermionic Hamiltonian into a qubit Hamiltonian $\mathcal{H}_\mathcal{Q}$, which is a weighted sum of Pauli strings as shown in the middle of Figure~\ref{fig:simulation-overview}.
The qubit Hamiltonian can then be compiled into circuits to simulate the original Fermionic system~\cite{trotter1959product,li2022paulihedral}.
Note that Fermion-to-qubit mapping is unique, and different mappings will lead to significantly different circuit implementation costs even for the same input Fermionic system, as shown in Figure~\ref{fig:simulation-overview}.

It is naturally desired for people to have Fermion-to-qubit mappings that can minimize the simulation circuit implementation cost. However, existing Fermion-to-qubit mappings are not yet satisfying and have significant further optimization potential. These mappings can be classified into two major types. 
The first type of approach is the constructive method, such as Jordan-Wigner transformation~\cite{jordan1928uber}, Bravyi-Kitaev transformation~\cite{bravyi2002fermionic}, and ternary tree mapping~\cite{miller2023bonsai,jiang2020optimal}.
Although some can generate asymptotically optimal mapping, these methods do not consider the input Hamiltonian and always generate the Fermion-to-qubit mapping using the same operator pattern, leading to suboptimal actual mapping overhead.
The second type of approach is the exhaustive search method, such as Fermihedral~\cite{liu2024fermihedral}. 
Fermihedral formulates the Fermion-to-qubit mapping search into a Boolean satisfiability (SAT) problem and uses an SAT solver to find the optimal mapping with minimal Pauli weight in the final qubit Hamiltonian.
Although this approach can find better mappings by encoding the input Hamiltonian in the SAT instance, it has a scalability problem due to the SAT's intrinsic hardness and fails to accommodate large-size practical cases. 


In this paper, we aim to push the boundary of Fermion-to-qubit mapping forward and design a new compilation algorithm that can generate lower-cost mapping results while leveraging the input Hamiltonian information in a scalable manner.
Formulating a good input-adaptive Fermion-to-qubit mapping is a complicated constrained optimization problem since valid Fermion-to-qubit mappings must satisfy certain constraints, and the input Hamiltonions can have different patterns across various Fermionic systems.
Our key observation is that existing approaches do not fully exploit the potential of mathematical objects in this domain, including the variants in the data structures and the properties of Pauli algebra. 
In particular, we notice that the flexibility of the ternary tree data structure can be the key to our new algorithm.
Indeed, ternary trees with different structures can yield valid but different Fermion-to-qubit mappings.
It is possible to alter the ternary tree structure based on the input Fermionic Hamiltonian to yield low-cost Fermion-to-qubit mappings without exhaustive search.

To this end, we propose the Hamiltonian-Adaptive Ternary Tree (\name{}) framework that can compile optimized Fermion-to-qubit mappings tailored to the input Fermionic Hamiltonian. 
\textbf{First}, \name{} comes with a new bottom-up ternary tree generation algorithm that can incorporate the input Fermionic Hamiltonian information into the generated tree structure. Starting from the leaf nodes, our algorithm gradually adds parent nodes to construct the tree to maximize the cancellation during Pauli operator multiplication in the input Hamiltonian, thus minimizing the Pauli weight on each qubit.
\textbf{Second}, we adapt our algorithm with a careful node selection and operator pairing process during tree construction. By traversing down and up the constructed tree, we can ensure that the resulting mapping will map the vacuum state in the Fermionic system to the all-zero state in the qubit mapping. This retains the important \textit{vacuum state preservation} property for a Fermion-to-qubit mapping. 
\textbf{Third}, we further apply a caching strategy to the tree structure during construction to reduce the complexity of traversing up and down during operator pairing from worst $O(N)$ to $O(1)$ when mapping $N$-mode Fermionic systems. Our optimization decreases the total complexity of the algorithm from $O(N^4)$ to $O(N^3)$.


We evaluate \name{} extensively against existing construction and exhaustive search methods on various Fermionic system models. The results show that \name{} generates close-to-optimal mappings for small-size cases compared with the optimal mapping from Fermihedral~\cite{liu2024fermihedral} and even outperforms the approximately optimal solutions of Fermihedral. 
For larger-scale cases where Fermihedral fails, \name{} outperforms those constructive methods (Jordan-Wigner~\cite{jordan1928uber}, Bravyi-Kiteav~\cite{bravyi2002fermionic}, balanced ternary tree~\cite{jiang2020optimal}) with $\sim 15\%$ reduction in Pauli weight, $5\sim 20\%$ reduction in circuit depth, and $10\sim30\%$ reduction in $CNOT$ gate in the simulation quantum circuit. 
In addition, \name{} achieves the lowest bias and variance in the noisy simulation experiments.
On the real system study on IonQ Forte 1 quantum computer, \name{} achieves the smallest variance and the second best average result (the best average result comes from Fermihedral's optimal mapping).


The major contributions of this paper can be summarized as follows:

\begin{enumerate}
    \item We propose \name{}, a ternary tree-based Fermion-to-qubit mapping compilation framework to construct input-adaptive ternary tree Fermion-to-qubit mappings based on the Fermionic Hamiltonian in polynomial time.
    \item We further improve our tree construction algorithm and successfully implement the \textit{vacuum state preservation}, a desired property for Fermion-to-qubit mapping without increasing the search complexity or affecting the generated mapping performance.
    \item By designing a map to cache the structure of the constructed tree, we reduce the overall complexity of our overall algorithm from $O(N^4)$ to $O(N^3)$ ($N$ is system size), achieving significant speed up for large cases.
    \item Evaluation on several Fermionic systems shows \name{} has much better scalability than exhaustive search and can outperform existing constructive methods on system-dependent performance, including circuit depth, gate count, and higher accuracy in noisy simulation and real-system experiments. 
\end{enumerate}

\section{Background}

In this section, we introduce the essential background for understanding this paper, including \textit{Fermionic systems},  \textit{Pauli string}, \textit{Fermion-to-qubit mapping}. 
More fundamental concepts of quantum mechanics and quantum computing can be found in~\cite{sakurai2020modern,nielsen2010quantum}.


\subsection{Fermionic System}\label{sec:fermionic-system}


Generally speaking, a Fermionic system contains a set of \textit{Fermionic modes}. Each $i^{th}$ Fermionic mode is captured by a $2$D Hilbert space called the \textit{Fock space} $\mathcal{F}_i$, and a pair of creation and annihilation operators $(a_i^\dagger, a_i)$.


\subsubsection{Fock Space}

According to the \textit{Pauli exclusion principle}~\cite{pauli1925uber}, a Fermionic mode has two exclusive states: unoccupied $\qty(\ket{0}_\mathcal{F})$ or occupied by one Fermion $\qty(\ket{1}_\mathcal{F})$. Its \textit{Fock space} $\mathcal{F}$ is the $2$D Hilbert space $\mathrm{span}\qty{\ket{0}_\mathcal{F},\ket{1}_\mathcal{F}}$, where $\{\ket{0}_\mathcal{F},\ket{1}_\mathcal{F}\}$ is called the \textit{Fock basis}.

The Fock space of a $N$-mode Fermionic system is a $2^N$D Hilbert space generated by the tensor products of the Fock space of all $N$ Fermionic modes. The Fock basis is thus the tensor products of each basis:
\begin{equation*}
    \ket{e_{N-1},\dots,e_0}_\mathcal{F}=\bigotimes_{i=0}^{N-1}\ket{e_i}_\mathcal{F}\quad e_i=0\text{ or }1
\end{equation*}
where $\ket{e_i}_\mathcal{F}$ corresponds to the $i^{th}$ Fermionic mode.

\subsubsection{Creation and Annihilation Operator}

Each Fermionic mode defines a pair of creation and annihilation operators $(a^\dagger_i, a_i)$. They act by increasing or decreasing the occupation number:
\begin{gather*}
    a^\dagger_i\ket{0_i}_\mathcal{F}=\ket{1_i}_\mathcal{F}\quad a_i\ket{1_i}_\mathcal{F}=\ket{0_i}_\mathcal{F} \\
    a^\dagger_i\ket{1_i}_\mathcal{F}=a_i\ket{0_i}_\mathcal{F}=\mathbf{0}
\end{gather*}
Notice that any annihilation operator $a_i$ that applies on the vacuum state $\ket{0\dots0}_\mathcal{F}$ produces $\mathbf{0}$: $a_i\ket{0\dots0}_\mathcal{F}\equiv\mathbf{0}$.

\subsubsection{Fermionic Hamiltonian}

Finally, the system's Hamiltonian characterizes the time evolution behavior of a Fermionic system. With creation and annihilation operators, we can second quantize~\cite{dirac1927quantum} the Hamiltonian of a Fermionic System by expressing it with the weighted sum of operator products. A simple 2-mode Fermionic system can have the following Hamiltonian:
\begin{equation}\label{eqn:example-hamiltonian}
    \mathcal{H}_\mathcal{F}=c_0a^\dagger_0 a_0+c_1a^\dagger_1a_1+c_2a_0^\dagger a_1^\dagger a_0a_1
\end{equation}


\subsection{Pauli String}\label{sec:pauli-string}

\textit{Pauli strings} are the most basic concepts to describe a qubit system, as all $N$-length Pauli strings form an orthogonal basis for $N$-qubit Hermitians~\cite{nielsen2010quantum}. A Pauli string $S$ of a $N$-qubit system is defined as the tensor product of $N$ Pauli operators:
\begin{equation*}
    S=\sigma_{N-1}\otimes\dots\otimes\sigma_0\quad\text{where }\sigma_i\in\{I,X,Y,Z\}
\end{equation*}
We say it has a length $N$. Each Pauli operator $\sigma_i$ operates solely on the qubit $q_i$. $I$ is the identity operator, and $\{X,Y,Z\}$ are the Pauli matrices:
\begin{equation*}
    X=\mqty[0 & 1 \\1 & 0],\ 
    Y=i\mqty[0 & -1 \\ 1 & 0],\ 
    Z=\mqty[1 & 0 \\ 0 & -1]
\end{equation*}


\subsubsection{Format}

A Pauli string could be written in two forms: 
\begin{itemize}
    \item \textit{$N$-length string}: the string lists operators from $\sigma_{N-1}$ to $\sigma_0$, such as $XYIZ=X\otimes Y\otimes I\otimes Z$
    \item \textit{Compact form}: identity operators ($I$) are omitted, and each non-identity Pauli operator is subscripted with the qubit it operates, such as $XYIZ=X_3Y_2Z_0$. Since operators are annotated, their order does not matter.
\end{itemize}


\subsubsection{Time Evolution Operator}\label{sec:time-evolution}

The ultimate goal of quantum simulation on a digital quantum computer is to execute the time evolution operator $e^{-i\mathcal{H}_\mathcal{Q}t}$ of a given qubit Hamiltonian $\mathcal{H}_\mathcal{Q}$ with quantum circuits. It begins with decomposing $\mathcal{H}_\mathcal{Q}$ to a unique linear combination of $N$-length Pauli strings:
\begin{equation*}
    \mathcal{H}_\mathcal{Q}=\sum c_jS_j\quad c_j\in\mathbb{R}
\end{equation*}

The time evolution operator $e^{-i\mathcal{H_Q}t}$ is then approximated by Trotterization and number of time steps $n$:
\begin{equation*}
    e^{-i\mathcal{H_Q}t}=e^{-it\sum c_jS_j}\approx\underbrace{\qty(\prod_je^{itc_jS_j/n})^n}_{\text{\textcircled{1}}}+\underbrace{O\qty(t^2/n)}_{\text{residue}}
\end{equation*}

\begin{figure}
    \centering
    \includegraphics[width=0.9\linewidth]{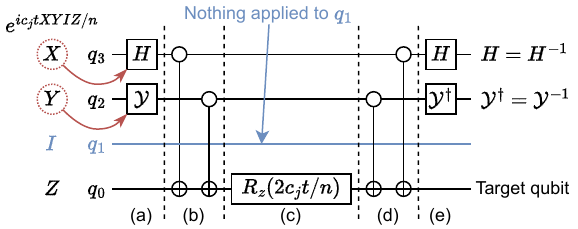}
    \caption{Circuit snippet of operator $\exp(itc_j XYIZ/n)$. $q_0$ is selected as the target qubit.}
    \label{fig:pauli-to-circuit}
\end{figure}

\textcircled{1} easily converts into a sequence of circuit snippets, each implementing a term $\exp(itc_jS_j/n)$ with basic gates. Figure~\ref{fig:pauli-to-circuit} (a)$\sim$(e) shows how term $\exp(itc_j XYIZ/n)$ is converted: (a) \revise{Apply $\mathcal{Y}=\frac{1}{\sqrt{2}}\mqty(1 & i \\ i & 1)$} gates to qubits that have $Y$ operator in the Pauli string ($q_2$), and apply $H$ to the ones have $X$ ($q_3$). \revise{This shifts the basis of the corresponding qubit from $Z$ into $Y$ or $X$.} (b) Use $CNOT$ gates to entangle all qubits with non-identity Pauli operators ($q_3,q_2$) to a target qubit ($q_0$). (c) Rotate the target qubit ($q_0$) with $R_z(2c_jt/n)$. (d)$\sim$(e) Apply inverse gates in (b) and (a).


\subsubsection{Pauli Weight}

The Pauli weight of a Pauli string $w(S)$ is the number of non-identity operators in it. For example, the string $XYIZ$ has weight $3$. The Pauli weight of a qubit Hamiltonian $\mathcal{H}_\mathcal{Q}=\sum c_iS_i$ is defined as the sum of weights of all decomposed Pauli strings: $w(\mathcal{H}_\mathcal{Q})=\sum w(S_i)$.
As discussed in Section~\ref{sec:time-evolution}, the identity operators in a Pauli string do not generate any gates. Thus, the Pauli weight of a qubit Hamiltonian is approximately proportional to the number of gates in the circuit of the time evolution operator $e^{-i\mathcal{H_Q}t}$.

\subsection{Fermion-to-Qubit Mapping}\label{sec:fermion-to-qubit-mapping}

To simulate a Fermionic system on a digital quantum computer, we have to bridge the Fermionic Hamiltonian $\mathcal{H}_\mathcal{F}$ to the qubit Hamiltonian $\mathcal{H}_\mathcal{Q}$. To do this, we use \textit{Fermion-to-qubit mapping} to find a set of Pauli strings to represent the creation and annihilation operators while satisfying the Fermionic anticommutation relationship. The mapping first pairs the $N$ creation and $N$ annihilation operators into $2N$ Majorana operators $M_j$:
\begin{equation}
    a^\dagger_j=\frac{M_{2j}-iM_{2j+1}}{2}\quad a_j=\frac{M_{2j}+iM_{2j+1}}{2}
\end{equation}
Each Majorana operator $M_j$ maps to a $N$-length Pauli string $S_j$, thus mapping the creation/annihilation operators and the Fermionic Hamiltonian to the qubit Hamiltonian.

For example, using the well-known Jordan-Wigner (\textsf{JW}) Fermion-to-qubit mapping~\cite{jordan1928uber}, the example Hamiltonian in Equation~\eqref{eqn:example-hamiltonian} is mapped to:
\begin{equation*}
    \begin{aligned}
        \mathcal{H}_\mathcal{F}\mapsto\mathcal{H}_\mathcal{Q}&=\frac{2c_0+2c_1-c_2}{4}II+\frac{c_2-2c_0}{4}IZ\\&+\frac{c_2-2c_1}{4}ZI-\frac{c_2}{4}ZZ
    \end{aligned}
\end{equation*}
with
\begin{equation*}
    \begin{aligned}
        & a_0^\dagger\mapsto0.5IX-0.5iIY & a_0\mapsto0.5IX+0.5iIY \\
        & a_1^\dagger\mapsto0.5XZ-0.5iYZ & a_1\mapsto0.5XZ+0.5iYZ
    \end{aligned}
\end{equation*}
and the Majorana operators are:
\begin{equation*}
    M_0=IX\quad M_1=IY\quad M_2=XZ\quad M_3=YZ
\end{equation*}


\section{Hamiltonian Adaptive Ternary Tree}
\label{sec:tree-construction}

We aim to design an efficient and scalable compilation framework to generate low-overhead Fermion-to-qubit mappings tailored to the input Hamiltonians. Our approach exploits the properties of Pauli strings and a unique data structure, the ternary tree, to yield Pauli strings satisfying the constraints of a valid Fermion-to-qubit. The Pauli strings are generated to maximize the Pauli operator cancellation during the multiplication of Majorana operators, leading to low Pauli weight post-mapping qubit Hamiltonians.



\subsection{Extracting Majorana Operators from Ternary Tree}
We first introduce the ternary tree and how to extract Majorana operators in the Fermion-to-qubit mapping from a ternary tree. 


\begin{figure}[t]
    \centering
    \includegraphics[width=\linewidth]{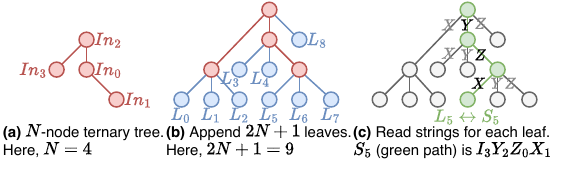}
    \caption{Example of a ternary tree and extracting Fermion-to-qubit mapping}
    \label{fig:ternary-tree-mapping}
\end{figure}

\subsubsection{Ternary Tree}

Ternary trees are a special type of tree. All nodes that are not leaves are denoted as internal nodes. Each internal node of a ternary tree has \textit{at most} three child nodes. A ternary tree is complete if every internal node has exactly three children. If a complete ternary tree has $n$ internal nodes, it has $2n+1$ leaves. Each path from the root to a leaf is always unique since a ternary tree does not have cycles.
For example, the four red nodes in Figure~\ref{fig:ternary-tree-mapping} (a) are the internal nodes of a ternary tree, and Figure~\ref{fig:ternary-tree-mapping} (b) shows a complete ternary tree after the leaves (blue nodes) are added to ensure each internal node has three child nodes.


\subsubsection{Extract Pauli Strings from Ternary Tree}

A ternary tree Fermion-to-qubit mapping for an $N$-mode Fermionic system can be constructed via $2N$ Pauli strings representing the $2N$ Majorana operators. The $2N$ Pauli strings are extracted from a complete ternary tree with $N$ internal nodes and $2N+1$ leaves. Each Pauli string is represented by a path from the root node to one leaf node in the ternary tree. This process is illustrated using the example in Figure~\ref{fig:ternary-tree-mapping}.

We denote each internal node by $In_j$ where $j=0,\dots, N-1$.
The internal node $In_j$ corresponds to the qubit $q_j$ in the qubit system to be mapped onto. The leaf nodes are denoted by $\{L_i\}$. In this example, the ternary tree has four internal nodes $In_0\sim In_3$ and nine leaves $L_0\sim L_8$.

Now, we take a path from the root node to one leaf. At each step in this path, there are three possible branches to select the next node because each internal node in a complete ternary tree has three children. The three branches are labeled by $X$, $Y$, and $Z$.
Then, a Pauli string can be determined by following the path. 
Suppose the Pauli string $S_i$ is constructed by the path from the root to the leaf node $L_i$, and the path is:
\begin{equation*}
    Path=In_a\text{(root)}\rightarrow In_b\rightarrow In_c\rightarrow\dots\rightarrow L_i
\end{equation*}
Each internal node $In_j$ contributes a Pauli operator in the Pauli string $S_i$. For the Pauli operator on qubit $q_j$, If $In_j\notin Path$, the operator is $I_j$. If $In_j\in Path$, based on the branch $In_j\rightarrow In_k$ it takes, the node contributes $X_j$, $Y_j$, or $Z_j$ if $In_k$ is the left, middle, or right child of $In_j$. $S_i$ equals the tensor product of all these operators to which the nodes contribute. For example, the green nodes in Figure~\ref{fig:ternary-tree-mapping} (c) represent a path from the root node to one leaf in the complete ternary tree. The path starts from node $In_2$ to node $In_0$ via the branch $Y$, so we have $Y_2$ in the Pauli string. Then the path goes from node $In_0$ to node $In_1$ via the branch $Z$, so we have $Z_0$. Similarly, we have $X_1$ because of the $X$ branch from $In_0$ to the leaf, and $I_3$ because $In_3$ is not in this path.
This Pauli string is $I_3Y_2X_1Z_0$.

As there are $2N+1$ leaves, paths from the root to these leaves can construct $2N+1$ distinct Pauli strings.
It can be proved that any $2N$ Pauli strings out of the $2N+1$ strings satisfy all the constraints required for the $2N$ Majorana operators in a valid Fermion-to-qubit mapping~\cite{miller2023bonsai,jiang2020optimal}. Thus, we can always construct a Fermion-to-qubit mapping for $N$ Fermionic modes through a complete ternary tree with $N$ internal nodes and $2N+1$ leaves.

\begin{figure}
    \centering
    \includegraphics[width=0.75\linewidth]{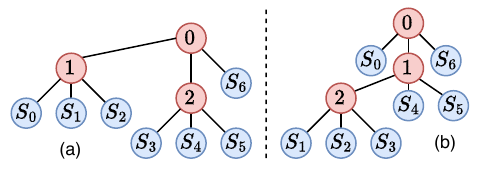}
    \caption{$3$-mode ternary tree Fermion-to-qubit examples. (a) Balanced ternary tree. (b) Unbalanced ternary tree.}
    \label{fig:btt-example}
\end{figure}

\subsection{Motivation: Drawback of Balanced Ternary Tree}\label{sec:balanced-ternary-tree}

Since the Pauli strings are determined by the paths from the root to the leaves in a ternary tree, the tree's structure will determine the generated Pauli strings and affect the overhead when simulating the corresponding qubit Hamiltonian.
The number of non-identity Pauli operators in one Pauli string is the number of internal nodes in this path. Therefore, to reduce the Pauli weight of the Pauli strings, previous work\cite{jiang2020optimal} employs the balanced ternary tree, which has the lowest depth on average. For $N$ Fermonic modes, it generates Pauli strings with an average Pauli weight of $\lceil\log_3(2N+1)\rceil$, which can be proved to be the theoretically optimal Pauli weight per Pauli string.


However, in this paper, we make a key observation that the minimal Pauli weight in the Pauli strings may not lead to the minimal circuit implementation overhead in the qubit Hamiltonian simulation. 
An actual Hamiltonian can be a complex expression of the Pauli strings. The Pauli operators may cancel with each other and become the identity during the multiplication of Pauli strings.
A motivation example is shown in Figure~\ref{fig:btt-example}.

Suppose we have a toy model $3$-mode Fermionic Hamiltonian (in Majorana operators) $\mathcal{H}_\mathcal{F}=c_1M_0M_5+c_2M_1M_3$. the mapping generated from a balanced ternary tree, as shown in Figure~\ref{fig:btt-example} (a), maps it to $\mathcal{H}_\mathcal{Q}=c_1(X_0X_1)(Y_0Z_2)+c_2(X_0Y_1)(Y_0X_2)=c_1'Z_0X_1Z_2+c_2'Z_0Y_1X_2$. The Pauli weight is 6. But the unbalanced ternary tree mapping, as shown in Figure~\ref{fig:btt-example} (b), can yield a lower Pauli weight. It maps the Hamiltonian to $\mathcal{H}_\mathcal{Q}=c_1(X_0)(Y_0Z_1)+c_2(Y_0X_1X_2)(Y_0X_1Z_2)=c_1''Z_0Z_1+c_2''Y_2$, which has a Pauli weight of 3.

The example above shows that a mapping generated by unbalanced ternary trees can have larger Pauli weights in the Pauli strings and smaller Pauli weights in the final qubit Hamiltonian because of the multiplication of the Pauli operators.
This motivates us to construct the ternary tree adaptively based on the input Fermionic Hamiltonian, which can further decrease the Pauli weight in the final qubit Hamiltonian.

\subsection{Hamiltonian Adaptive Ternary Tree Construction}\label{sec:naive-algorithm}

\begin{figure}[ht]
    \centering
    \includegraphics[width=\linewidth]{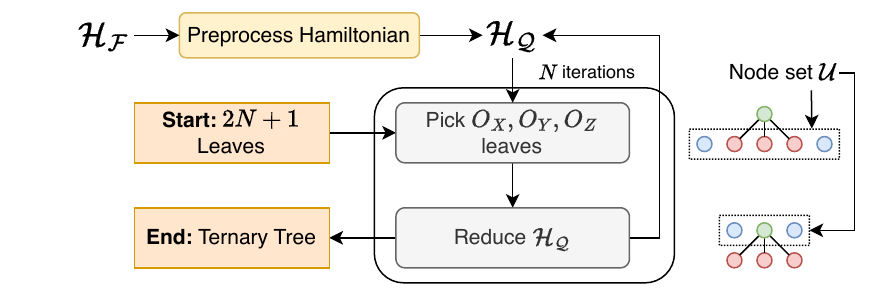}
    \caption{Algorithm for ternary tree construction}
    \label{fig:algorithm-process}
\end{figure}

This section introduces our ternary tree construction algorithm that can exploit cancellation between Majorana operators for an input Fermionic Hamiltonian. Suppose we have a $N$-mode Fermionic Hamiltonian. Our compilation takes a bottom-up approach, starting with all the $2N+1$ leaves. We gradually append parent nodes to the tree and finally reach the root. These parent nodes are always the internal nodes representing qubits. Our algorithm will search for a tree structure to lower the Pauli weight on each qubit. The pseudo-code is in Algorithm~\ref{alg:tree-construction}, and an algorithm overview is in Figure~\ref{fig:algorithm-process}. The algorithm details are explained below.

\begin{algorithm}[t]
\caption{Hamiltonian Adaptive Ternary Tree \revise{Construction}}\label{alg:tree-construction}
\begin{algorithmic}[1]
\Require $\mathcal{H}_\mathcal{F}$: Hamiltonian of the Fermionic system
\Ensure $\{S_i\}$: $2N$ anticommute Pauli strings generated by the ternary tree, used as $2N$ Majorana operators.

\State $\mathcal{H}_\mathcal{Q}\gets\mathtt{preprocess}(\mathcal{H}_\mathcal{F})$
\State $\mathcal{U}\gets\{O_0,\dots,O_{2N}\}$ \Comment{initial node set $\mathcal{U}$}
\For {$i$ \textbf{from} $0$ \textbf{to} $N$}
    \State $w\gets\infty$
    \State $selection\gets(\mathrm{null},\mathrm{null},\mathrm{null})$
    \For{$O_X,O_Y,O_Z\in\mathtt{permutation}(\mathcal{U}, 3)$}
        \State $w'\gets\mathtt{pauli\_weight}(\mathcal{H}_\mathcal{Q}, (O_X, O_Y, O_Z), i)$ 
        \Statex \Comment{Pauli weight on qubit $i$}
        \If {$w'<w$}
            \State $w\gets w'$
            \State $selection\gets(O_X, O_Y, O_Z)$
        \EndIf
    \EndFor

    \State $\mathcal{U}.\mathtt{remove}(O_X)$; $\mathcal{U}.\mathtt{remove}(O_Y)$; $\mathcal{U}.\mathtt{remove}(O_Z)$
    \State $\mathcal{U}.\mathtt{add}(O_{2N+1+i})$
    \State $O_{2N+1+i}.(X, Y, Z)\gets (O_X, O_Y, O_Z)$
    \State $\mathcal{H}_\mathcal{Q}\gets\mathtt{reduce}(\mathcal{H}_\mathcal{Q}, (O_X, O_Y, O_Z)\mapsto O_{2N+1+i})$
    \Statex \Comment{reduce Hamiltonian $\mathcal{H}_\mathcal{Q}$}
\EndFor
\State $O_{root}\gets \mathcal{U}.\mathtt{pop}()$
\State $\{S_i\}\gets\mathtt{walk\_tree}(O_{root})$
\State \textbf{return} $\{S_i\}$
\end{algorithmic}
\end{algorithm}

\subsubsection{Setup}

The algorithm compiles a complete ternary tree Fermion-to-qubit mapping for an $N$-mode input Fermionic Hamiltonian with $2N+1$ leaf nodes and $N$ internal nodes. 
For simplicity, all the nodes are denoted by $O_i$ in the rest of this paper.
The $2N+1$ leaf nodes are denoted by $O_0\sim O_{2N}$. The $N$ internal nodes are denoted by $O_{2N+1}\sim O_{3N}$. 
Leaf node $O_{0\leq \mathbf{i}\leq 2N}$ (previously $L_\mathbf{i}$ in the ternary tree example in Figure~\ref{fig:ternary-tree-mapping}) corresponds to the Pauli string $S_\mathbf{i}$ and is used as the Majorana operator $M_i$ in the Fermion-to-qubit mapping. Internal node $O_{2N+1\leq2N+1+\mathbf{i}\leq 3N}$ (previously $In_\mathbf{i}$ in Figure~\ref{fig:ternary-tree-mapping}) corresponds to qubit $q_\mathbf{i}$. The algorithm starts with $2N+1$ leaves in the ternary tree as the initial node set $\mathcal{U}=\{O_0,\dots, O_{2N}\}$.

The initial Fermionic Hamiltonian $\mathcal{H_F}$ is $\mathtt{preprocessed}$ to substitute all creation and annihilation operators with Majorana operators $\{M_i\}$, then $\mathcal{H_F}$ is mapped to a qubit Hamiltonian $\mathcal{H_Q}$ by using Pauli strings to represent Majorana operators: $M_i\rightarrow S_i$. The algorithm finds a concrete Pauli string for each $S_i$.
For example, we have a 3-mode Hamiltonian:
\begin{equation}\label{eqn:ternary-hamiltonian}
\begin{aligned}
     \mathcal{H}_\mathcal{F}= & \ a_0^\dagger a_0+2a_1^\dagger a_2^\dagger a_1a_2 \\ = & \ 0.5i\cdot M_0M_1-0.5i\cdot M_2M_3-0.5i\cdot M_4M_5 \\ & +0.5\cdot M_2M_3M_4M_5
\end{aligned}
\end{equation}
The algorithm begins with $\mathcal{U}=\{O_0,\dots,O_6\}$, each carries $S_0,\dots, S_6$.
The Hamiltonian is preprocessed to $\mathcal{H_Q}=0.5i\cdot S_0S_1-0.5i\cdot S_2S_3-0.5i\cdot S_4S_5+0.5\cdot S_2S_3S_4S_5$.

\subsubsection{Iteration}

The algorithm takes $N$ iterations, from $0$ to $N-1$. In the $i^{th}$ step, the algorithm picks three nodes from $\mathcal{U}$ and grows a parent $O_{2N+1+i}$ for them. The three nodes are removed from $\mathcal{U}$, and $O_{2N+1+i}$ is added to $\mathcal{U}$, which reduces the size of $\mathcal{U}$ by 2. This settles the operator on qubit $q_i$ for all strings. The three nodes are carefully selected to produce a minimum Hamiltonian Pauli weight on $q_i$. Details are discussed as follows:

\begin{itemize}
    \item \textbf{Node Selection} (line $6\sim12$ in Algorithm~\ref{alg:tree-construction}): Suppose we select three nodes $O_X$, $O_Y$, $O_Z$ as the $X, Y, Z$ child for node $O_{2N+1+i}$. Based on definition, Pauli strings $S_X, S_Y, S_Z$ correspond to nodes $O_X, O_Y, O_Z$ have $X, Y, Z$ operators on qubit $q_i$ while the rest always have $I$, as shown in Figure~\ref{fig:select-nodes}
    \begin{figure}[t]
        \centering
        \includegraphics[width=0.75\linewidth]{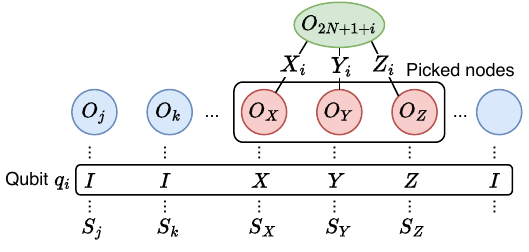}
        \caption{Picked nodes $O_X, O_Y, O_Z$, their parent $O_{2N+1+i}$ and corresponding Pauli operators on qubit $q_i$}
        \label{fig:select-nodes}
    \end{figure}
    
    For each possible selection of $O_X, O_Y, O_Z$, we calculate the Pauli weight for the Hamiltonian on qubit $q_i$ ($\mathtt{pauli\_weight}$ in Algorithm~\ref{alg:tree-construction} line 7). It can be done by only calculating the $i^{th}$ Pauli operator of the Hamiltonian. Based on the Pauli weight and \textit{greedy} strategy, $O_X, O_Y, O_Z$ should give the \textit{minimum Pauli weight} on qubit $q_i$ among all selections.
    
    For the example Hamiltonian in Equation~\eqref{eqn:ternary-hamiltonian}, we will pick $O_0, O_1, O_6$ in the first step. Only $S_0, S_1, S_6$ will have $X, Y, Z$ operator on qubit $q_0$ while the rest have $I$, \revise{as shown in Figure~\ref{fig:step3} \textbf{\textcircled{1}}}. The Pauli weight is:
    \begin{gather*}
        S_0S_1\mapsto XY= Z (1) \\
        S_2S_3,\ S_4S_5\mapsto II= I (0) \\
        S_2S_3S_4S_5\mapsto II= I(0) \\
        \text{\textbf{Total Pauli weight}}=1+0+0+0=1
    \end{gather*}

    \item \textbf{Update tree and node set $\mathcal{U}$} (line $13\sim15$ in Algorithm~\ref{alg:tree-construction}): Remove $O_X,O_Y,O_Z$ from $\mathcal{U}$ and add $O_{2N+1+i}$ to $\mathcal{U}$. Connect $O_X, O_Y, O_Z$ to the $X, Y, Z$ child of node $O_{2N+1+i}$. \revise{The node set $\mathcal{U}$ for the example is updated as shown in Figure~\ref{fig:step3} \textbf{\textcircled{2}}}.
    \begin{figure}[ht]
        \centering
        \includegraphics[width=\linewidth]{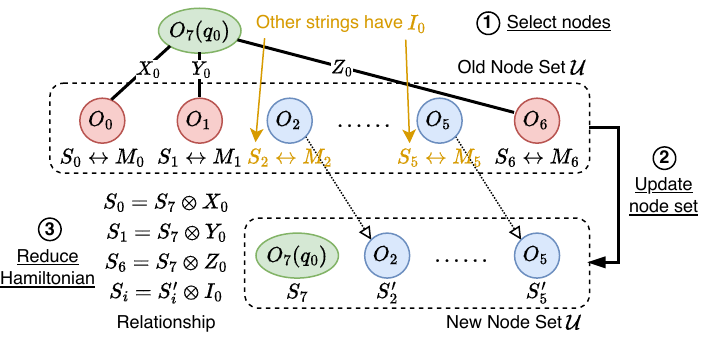}
        \caption{\revise{Example: \textbf{\textcircled{1}} select $O_0, O_1, O_6$ nodes and construct their parent node $O_7 (q_0)$. Thus, $S_0, S_1$ and $S_6$ has $X_0, Y_0$ and $Z_0$ in it. All the rest of Pauli strings get $I_0$;\textbf{\textcircled{2}} update node set $\mathcal{U}$ by removing $O_0, O_1, O_6$ and inserting $O_7$; \textbf{\textcircled{3}} reduce Hamiltonian $\mathcal{H}_\mathcal{Q}$ by the relationship: $O_7$ carries $S_7$, which is the \textit{common part} of $S_0, S_1$ and $S_6$, by $\{S_0,S_1,S_6\}=S_7\otimes\{X,Y,Z\}$. All other nodes carried another Pauli string $S_i'$ such that $S_i=S_i'\otimes I_0$.}}
        \label{fig:step3}
    \end{figure}

    \item \textbf{Reduce Hamiltonian} ($\mathtt{reduce}$ in Algorithm~\ref{alg:tree-construction} line 16): The node $O_{2N+1+i}$ carries a Pauli string $S_{2N+1+i}$ such that:
    \begin{equation*}
        S_{\{X,Y,Z\}}=S_{2N+1+i}\otimes \{X,Y,Z\}_i
    \end{equation*}
    For the rest of the nodes $\forall O_j\notin\{O_X, O_Y, O_Z\}$, we can also find a Pauli string $S_j'$ such that $S_j=S_j'\otimes I_i$. We update $O_j$ to carry $S_j'$ instead of $S_j$.
    
    Then, we subsititue $S_j$ with $S_j'$ and $\{S_X, S_Y, S_Z\}$ with $S_{2N+1+i}$ in $\mathcal{H_Q}$, \revise{and remove operators on qubit $q_i$ from $\mathcal{H}_\mathcal{Q}$ since it is settled for all Pauli strings and will not affect total Pauli weight.}

    In the example, we will introduce node $O_7$ representing qubit $q_0$ that carries $S_7$. \revise{The relationships between $S_7$ and $S_0,S_1,S_6$ and between $S_i'$ and $S_i$ are also shown in Figure~\ref{fig:step3} \textbf{\textcircled{3}}. By substituting using the relationship:}
    \revise{\begin{gather*}
        S_0\mapsto S_7\otimes X_0\quad S_1\mapsto S_7\otimes Y_0\quad S_6\mapsto S_7\otimes Z_0 \\
        S_i\mapsto S_i'\otimes I_0
    \end{gather*}}
    \revise{each term of Hamiltonian $\mathcal{H}_\mathcal{Q}$ in Equation (\ref{eqn:ternary-hamiltonian}) reduces to:}
    \revise{\begin{gather*}
        S_0S_1\mapsto S_7S_7=I\otimes Z_0 \\
        S_2S_3\mapsto S_2'S_3'\otimes I_0,\ S_4S_5\mapsto S_4'S_5'\otimes I_0\\
        S_2S_3S_4S_5\mapsto S_2'S_3'S_4'S_5'\otimes I_0
    \end{gather*}}
    \revise{Operator on $q_0$ ($\dots\otimes\sigma_0$ part) is settled and can be removed since it does not further affect the total Pauli weight. $\mathcal{H}_\mathcal{Q}$ reduces to:}
    \revise{\begin{equation*}
        \mathcal{H}_\mathcal{Q}'=0.5i\cdot S_2'S_3'-0.5i\cdot S_4'S_5'+0.5\cdot S_2'S_3'S_4'S_5'
    \end{equation*}}

    In the next iteration step, we have node set $\mathcal{U}=\{O_2, O_3, O_4, O_5, O_7\}$ and the reduced Hamiltonian $\mathcal{H}_\mathcal{Q}'$. \revise{We can then repeat the above steps with the node set $\mathcal{U}$ and $\mathcal{H}_\mathcal{Q}'$ until all $N=3$ iterations are performed.}
\end{itemize}

\subsubsection{Termination}

The algorithm always terminates after $N$ iterations since the initial size of $\mathcal{U}$ is $2N+1$ and reduces by two in each step. The final only node in $\mathcal{U}$ is the root of the ternary tree.

\subsubsection{Complexity}

We disregard the complexity of calculating the Pauli weight in each iteration step since it is input-dependent. We estimate its overhead to be a constant determined by the input Hamiltonian.
In the algorithm, we compared all possible selections of $O_X, O_Y, O_Z$, which are $\binom{N}{3}\sim N^3$ choices. Thus, the complexity of each step is $O(N^3)$. For total $N$ steps, the complexity is $O(N^4)$.

\section{\revise{Vacuum} State Preservation and Performance Optimization}
\label{sec:optimized-algorithm}

The algorithm described in Section~\ref{sec:naive-algorithm} decreases the Hamiltonian Pauli weight but fails to retain one desired property of a Fermion-to-qubit mapping: \textit{vacuum state preservation}. In addition, the computation complexity $O(N^4)$ is still high. In this section, we further optimize our algorithm to a) ensure \textit{vacuum state preservation} in the generated mapping and b) reduce its complexity to $O(N^3)$.

\subsection{Vacuum State Preservation}

In Fermion-to-qubit mappings, it is desired to map the vacuum state of a Fermionic system $\ket{0,\dots,0}_\mathcal{F}$ to the zero qubit state $\ket{0}^{\otimes N}$ of the qubits. It is called \textit{Vacuum State Preservation}, allowing lower overhead for state preparation in quantum simulation. 
This property is achieved by ensuring that the annihilation operators always produce $\mathbf{0}$ when applying on the vacuum state:
\begin{equation*}
\forall j, a_j\ket{0,\dots,0}_\mathcal{F}=\mathbf{0}\Leftrightarrow\frac{M_{2j}+iM_{2j+1}}{2}\ket{0}^{\otimes N}=\mathbf{0}
\end{equation*}
To satisfy the right-hand-side equation, the corresponding Pauli strings $S_{2j}, S_{2j+1}$ of Majorana operators $M_{2j}, M_{2j+1}$ should have a $(X, Y)$ Pauli operator pair on a qubit, since $((X+iY)/2)\ket{0}\equiv\mathbf{0}$, and all the rest operators act the same on $\ket{0}$, i.e., the $k^{th}$ Pauli operator $\sigma_k^{2j},\sigma_k^{2j+1}$ of $S_{2j},S_{2j+1}$ satisfies $\sigma_k^{2j}\ket{0}=\sigma_k^{2j+1}\ket{0}$~\cite{miller2023bonsai}. In the vanilla balance ternary tree~\cite{miller2023bonsai,jiang2020optimal}, this is achieved by re-assigning the Pauli strings to Majorana operators.


\subsection{Operator Pairing In Tree Construction}\label{sec:operator-pairing}

However, we cannot re-assign Pauli strings to Majorana operators in our Hamiltonian-adaptive ternary tree construction because we have already assumed the Pauli string $S_j$ is assigned to the Majorana operator $M_j$. Reassignment destroys the Pauli operator cancellation created in our algorithm. 

Instead of re-assigning Pauli strings, we improve our tree construction algorithm by enforcing \textit{vacuum state preservation} during the node selection. That is, we only freely select $O_X$ and $O_Z$, and $O_Y$ is determined based on $O_X$ and $O_Z$. With careful construction leveraging the property of ternary trees, we can guarantee that all the Majorana operator pairs $(M_{2j}, M_{2j+1})$ have an $(X, Y)$ Pauli operator pair on one qubit, and operators on other qubits act the same on $\ket{0}$, and thus ensure vacuum state preservation.

We first define two new concepts in our new algorithm:
\begin{enumerate}[I)]
    \item \textit{\textit{$Z$-descendant}}: The \textit{$Z$-descendant} of a node $O$, denoted as $desc_Z(O)$, is defined as the \textit{rightmost leaf} of the subtree with root $O$. It can be reached by traversing down all the $Z$ branches. If $O$ is a leaf, then $desc_Z(O)=O$.
    \item \textit{Valid Pair}: Two Pauli strings $(S_{2j}, S_{2j+1})$ forms a \textit{valid pair} if they share a $(X, Y)$ pair on a qubit and all other operators act the same on $\ket{0}$. If we have $(S_{2j}, S_{2j+1})$ is a \textit{valid pair} forall $0\leq j<N$, then \textit{vacuum state preservation} is guranteed.
\end{enumerate}

\begin{figure}[ht]
    \centering
    \includegraphics[width=0.75\linewidth]{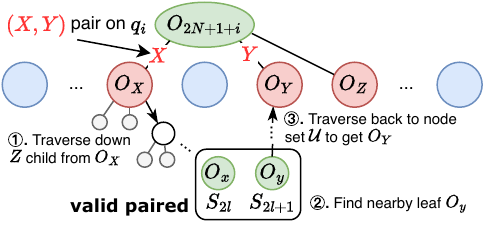}
    \caption{Procedure of finding $O_Y$ based on $O_X, O_Z$}
    \label{fig:optimized-algorithm}
\end{figure}

We now introduce our improved tree construction algorithm~\ref{alg:optimized-algorithm}. The overall structure is similar to Algorithm~\ref{alg:tree-construction}, but node selection is changed.
In each step $i$, as shown in Algorithm~\ref{alg:optimized-algorithm} instead of selecting all $O_X, O_Y, O_Z$ nodes, we now only select nodes $O_X, O_Z$ as the $X, Z$ children of node $O_{2N+1+i}$ and find an appropriate $O_Y$ to ensure exists $l$, such that $desc_Z(O_X)$ corresponds to $S_{2l}$ and \revise{$desc_Z(O_Y)$} corresponds to $S_{2l+1}$. Thus, $(S_{2l}, S_{2l+1})$ is valid paired since they have a $(X, Y)$ pair on qubit $q_i$, and for all $k\neq i$, the $k^{th}$ operators for $S_{2l},S_{2l+1}$ are either the same or a $Z,I$ pair, whereas $Z\ket{0}\equiv I\ket{0}$, as shown in Figure~\ref{fig:optimized-algorithm}.
\begin{algorithm}[t]
\caption{Optimized \name}\label{alg:optimized-algorithm}
\begin{algorithmic}[1]
\Require $\mathcal{H}_\mathcal{F}$: Hamiltonian of the Fermionic system
\Ensure $\{S_i\}$: $2N$ Majorana operators generated by the ternary tree
\State $\cdots$ \Comment{Algorithm~\ref{alg:tree-construction}, line $1\sim 2$}
\For {$i$ \textbf{from} $0$ \textbf{to} $N$}
    \State $\cdots$ \Comment{Algorithm~\ref{alg:tree-construction}, line $4\sim 5$}
    \For {$O_X,O_Z\in\mathtt{permutation}(\mathcal{U}, 2)$}
        \State $O_x\gets\mathtt{desc_Z}(O_X)$
        \If {$x = 2N$}
            \State \textbf{continue} \Comment{$O_x$ is the last operator, no pairing}
        \EndIf
        
        \If {$x$ \textbf{is} even}
            \State $O_y\gets O_{x+1}$
            \State $O_Y\gets\mathtt{traverse\_up}(O_y, \mathcal{U})$
        \Else
            \State $O_y\gets O_{x-1}$
            \State $O_Y\gets\mathtt{traverse\_up}(O_y, \mathcal{U})$
            \State $\mathtt{swap}(O_X, O_Y)$ \Comment{swap $Y, X$ to $X, Y$}
        \EndIf
        \State $\cdots$ \Comment{Algorithm~\ref{alg:tree-construction}, line $7\sim 11$}
    \EndFor
    \State $\cdots$ \Comment{Algorithm~\ref{alg:tree-construction}, line $13\sim 16$}
\EndFor
\State $\cdots$ \Comment{Algorithm~\ref{alg:tree-construction}, line $18\sim 20$}
\Statex
\end{algorithmic}
\begin{algorithmic}[1]
    \Procedure{$\mathtt{desc_Z}$}{$O$}
    \While {$O$ \textbf{is not} leaf}
        \State $O\gets O.Z$
    \EndWhile
    \State \textbf{return} $O$
\EndProcedure
\Statex
\end{algorithmic}
\begin{algorithmic}[1]
    \Procedure{$\mathtt{traverse\_up}$}{$O, \mathcal{U}$}
    \While {$O\notin \mathcal{U}$}
        \State $O\gets O.\text{parent}$
    \EndWhile
    \State \textbf{return} $O$
\EndProcedure
\end{algorithmic}
\end{algorithm}

Suppose $O_X$ and $O_Z$ are selected. We traverse down along the $Z$ child from node $O_X$ until reaching a leaf of the final leaf $O_{x}=desc_Z(O_X)$ (where $0\leq x<2N+1$), as shown in Figure~\ref{fig:optimized-algorithm} \textcircled{1}. Then, we select its nearby leaf $O_y$ as $desc_Z(O_Y)$ based on $O_x$ to ensure the corresponding Pauli strings $S_x, S_y$ form a valid pair $(S_{x=2l},S_{y=2l+1})$, as shown in Figure~\ref{fig:optimized-algorithm} \textcircled{2}:
\begin{itemize}\label{sec:locate-Oy}
    \item If $x=2N$, then $O_x$ is the rightmost leaf. Discard this selection and pick $O_X, O_Z$ again ($S_{2N}$ is always discarded in the ternary tree mapping and does not pair).
    \item If $x$ is \textit{even} ($x=2l$), let $y=x+1$ ($y=2l+1$).
    \item If $x$ is \textit{odd} ($x=2l+1$), let $y=x-1$ ($y=2l$). We must also exchange $O_X$ with $O_Y$ when we get $O_Y$ (Line 15, Algorithm~\ref{alg:optimized-algorithm}) to ensure $(S_x, S_y)$ has a $(X, Y)$ Pauli operator pair, instead of $(Y, X)$. 
\end{itemize}

To find $O_Y$ based on $O_y=desc_Z(O_Y)$, we can traverse up from $O_y$ until we reach a node $O_Y$ in the current node set $\mathcal{U}$, as shown in Figure~\ref{fig:optimized-algorithm} \textcircled{3}. $O_Y$ is the ancestor of $O_y$, as it is the root of the subtree in which $O_y$ is located.

Finally, $O_Y$ is chosen as the $Y$ child of $O_{2N+1+i}$ to ensure valid string pairing. For each selection of $O_X, O_Z$ in step $i$, we calculate the Pauli weight similarly to the original algorithm and construct the node based on the selection of $O_X, O_Y, O_Z$ that gives a minimum Pauli weight on qubit $q_i$, then reduce the Hamiltonian similarly to the original algorithm. 
 
Consider our previous example, where $\mathcal{H}_\mathcal{F}=a_0^\dagger a_0+2a_1^\dagger a_2^\dagger a_1a_2=0.5i\cdot M_0M_1-0.5i\cdot M_2M_3-0.5i\cdot M_4M_5+0.5\cdot M_2M_3M_4M_5$. We have already selected $O_0, O_1, O_6$ as the children of $O_7$ in the first step, and the Hamiltonian is reduced to $\mathcal{H_Q}=0.5iS_2'S_3'-0.5iS_4'S_5'+0.5S_2'S_3'S_4'S_5'$.

\begin{figure}[ht]
    \centering
    \includegraphics[width=0.95\linewidth]{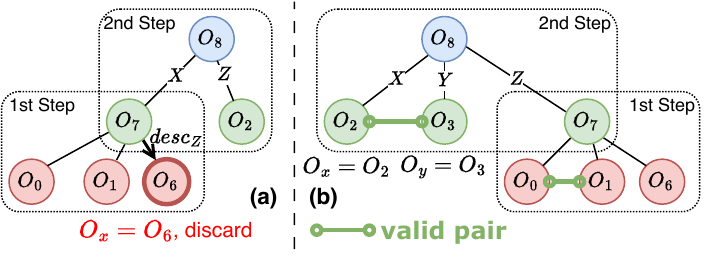}
    \caption{\revise{An example of operator pairing in tree construction. In (a), we select $O_7, O_2$ as $O_X, O_Z$, but $O_x=desc_Z(O_7)=O_6$ is the rightmost leaf, thus we discard this selection. In (b), we select $O_2, O_7$ as $O_X, O_Z$ and found $O_3$ as $O_Y$. Green lines show the valid pair formed by Pauli strings.}}
    \label{fig:optimized-example}
\end{figure}

In the second step, we first choose $O_7, O_2$ as $O_X, O_Z$, as shown in \revise{Figure~\ref{fig:optimized-example} (a)}. However,  $O_x=desc_Z(O_7)=O_6$ is the rightmost operator, so we discard this selection and move on. Then, we choose $O_2, O_7$ as $O_X, O_Z$, as shown in \revise{Figure~\ref{fig:optimized-example} (b)}. Here, $O_x=desc_Z(O_2)=O_2$, thus $O_y=O_{2+1}=O_3$ and traverse back to $S$ gives $O_Y=O_3$. This choice also minimizes the Pauli weight. $O_2,O_3,O_7$ are the children of $O_8$.

We then follow the original procedure and reduce the Hamiltonian:
\begin{gather*}
    S'_2S'_3\mapsto XY \rightarrow Z (1) \\ 
    S'_4S'_5\mapsto II\rightarrow I (0) \\
    S'_2S'_3S'_4S'_5\mapsto XYII\rightarrow Z(1) \\
    \text{\textbf{Total Pauli weight}}=1+0+1=2
\end{gather*}

We can check that Majorana operators are paired validally. $(M_2, M_3)\mapsto(S_2, S_3)$ have a $(X, Y)$ pair on qubit $1$, and $(M_0, M_1)\mapsto(S_0, S_1)$ have a $(X, Y)$ pair on qubit $q_0$.
 
\subsection{Optimizing Operator Pairing}\label{sec:optimized-pairing}

In the algorithm above, the complexity at the node selection is reduced because we only select two operators $O_X, O_Z$ in each step. The computation complexity is $\binom{N}{2}\sim O(N^2)$. 
However, it introduces new overhead traversing to $O_x$ and up to $O_Y$. In the worst case, the complexity of traversing up and down can be $O(i)$, leading to a total complexity of $O(N^4)$ (there are $N$ steps).

In this section, we further optimize the algorithm by reducing the traversing time from $O(N)$ to $O(1)$.
This improvement is made based on the following \textit{statement}:

\begin{mdframed}
In step $i$, the $O_y$ we find and its ancestor $O_Y\in \mathcal{U}$ must always satisfy $O_y=desc_Z(O_Y)$
\end{mdframed}

\textit{Proof}: We first prove two lemmas.
\begin{mdframed}
\textbf{Lemma 1}: Before step $i$, a leaf $O$ is not paired with another leaf if and only if it is a Z descendant of a node in $\mathcal{U}$. In other words, $\exists O'\in \mathcal{U}, O=desc_Z(O')$.

\textit{Proof}: We prove it by induction.
\begin{enumerate}[(I)]
    \item Before step $i=0$, $\mathcal{U}$ includes all leaves. They are all unpaired and Z descendants of themselves.
    \item Suppose Lemma 1 is satisfied before step $i$. We prove it holds after step $i$ (holds before step $i+1$). 
    
    After step $i$, we find $O_X$, $O_Y$, and $O_Z$ as the $X, Y$, and $Z$ children of node $O_{2N+1+i}$. Based on the induction hypothesis, all unpaired leaves in the $O_X, O_Y$, and $O_Z$ subtree are $O_x=desc_Z(O_X), O_y=desc_Z(O_Y)$, and $O_z=desc_Z(O_Z)$. Our algorithm pairs $O_x$ with $O_y$ while leaving $O_z$ unpaired. Since $O_Z$ is the $Z$ child of $O_{2N+1+i}$, we have $O_z=desc_Z(O_Z)=desc_Z(O_{2N+1+i}\in \mathcal{U})$ and it is the only unpaired leaf in the $O_{2N+1+i}$ subtree. 
    
    Now, we can check the lemma. $O_x$ and $O_y$ are paired and no longer Z descendants since their path to $O_{2N+1+i}$ takes $X$ or $Y$ branch. $O_z$ is unpaired and still Z descendant of $O_{2N+1+i}\in S$. The lemma holds on $O_x, O_y$, and $O_z$. All other leaves are untouched, and the lemma holds on them. \hfill$\square$
\end{enumerate}
\end{mdframed}

\begin{mdframed}
\textbf{Lemma 2}: $O_y$ is not paired in step $i$.

\textit{Proof}: Notice that $\forall 0\leq l<N$, leaf $O_{2l}$ must be paired with $O_{2l+1}$. $O_x$ is not paired in step $i$ due to Lemma 1. Then, based on how we find $O_y$ in Section~\ref{sec:locate-Oy}, $O_y$ must be unpaired. Otherwise, it leads to a contradiction. \hfill$\square$
\end{mdframed}
Based on Lemma 1 and $O_x=desc_Z(O_X)$, we have $O_x$ must be unpaired. Then, based on Lemma 2, $O_y$ is also unpaired. Again, by Lemma 1, $O_y$ is a Z descendant, indicating $O_y=desc_Z(O_Y)$.

The observation hints that the traversing up and down procedure ($\mathtt{desc_Z}$ and $\mathtt{traverse\_up}$ in Algorithm~\ref{alg:optimized-algorithm}) only involves the $Z$ descendants and their ancestors. Thus, we could keep two maps: $O\mapsto desc_Z(O)$ and $desc_Z(O)\mapsto O$ to reduce the complexity of traversing from $O(N)$ to $O(1)$. The map updates when constructing a new parent in each step, as shown in Algorithm~\ref{alg:update-map}. Overall, this optimization reduced the total complexity from $O(N^4)$ to $O(N^3)$.

\begin{algorithm}
\caption{Optimized Algorithm~\ref{alg:optimized-algorithm} with $O\mapsto desc_Z(O)$ and $desc_Z(O)\mapsto O$ maps}\label{alg:update-map}
\begin{algorithmic}[1]
\Require $m_{down}$: $O\mapsto desc_Z(O)$ map

$m_{up}$: $desc_Z(O)\mapsto O$ map

\For {$i$ \textbf{from} $0$ \textbf{to} $2N$} \Comment{initialize maps}
    \State $m_{down}[O_i]\gets O_i$
    \State $m_{up}[O_i]\gets O_i$
\EndFor
\State $\cdots$ \Comment{Algorithm~\ref{alg:tree-construction}, line $1\sim 2$}
\For {$i$ \textbf{from} $0$ \textbf{to} $N$}
    \State $\cdots$ \Comment{Algorithm~\ref{alg:optimized-algorithm}, line $3\sim 19$}
    \State $O_{2N+1+i}.(X, Y, Z)\gets (O_X, O_Y, O_Z)$

    \State $Zdesc\gets m_{down}[O_Z]$
    \State $m_{down}[O_{2N+1+i}]\gets Zdesc$ 
    \State $m_{up}[Zdesc]\gets O_{2N+1+i}$
    \State $\cdots$ \Comment{Algorithm~\ref{alg:tree-construction}, line $16$}
\EndFor
\State $\cdots$ \Comment{same as Algorithm~\ref{alg:tree-construction}, line $18\sim 20$}
\Statex
\end{algorithmic}
\begin{algorithmic}[1]
    \Procedure{$\mathtt{desc_Z}$}{$O$} \Comment{$O(1)$ time}
    \State \textbf{return} $m_{down}[O]$
\EndProcedure
\Statex
\end{algorithmic}
\begin{algorithmic}[1]
    \Procedure{$\mathtt{traverse\_up}$}{$O, \mathcal{U}$} \Comment{$O(1)$ time}
    \State \textbf{return} $m_{up}[O]$
\EndProcedure
\end{algorithmic}
\end{algorithm}



\begin{table*}[ht]
\caption{Evaluation result of Electronic Structure Model. The best results of each metric are highlighted in bold. `--' indicates the case is too large to solve by Fermihedral (\textsf{FH}). `*' indicates that Fermihedral only finds an approximately optimal solution. \revise{'frz' indicates freeze core transformation. 'sto3g'~\cite{stewart1970small} indicates the basis.}}
\centering
\label{tab:electronic-strcuture}
\resizebox{\linewidth}{!}{%
\begin{tabular}{|l|l||lllll||lllll||lllll|}
\hline
\multicolumn{1}{|c|}{\multirow{2}{*}{\textbf{Molecule}}} & \multicolumn{1}{c||}{\multirow{2}{*}{\textbf{Modes}}} & \multicolumn{5}{c||}{\textbf{Pauli Weight}}                                                                                                                                                               & \multicolumn{5}{c||}{\textbf{$CNOT$ Gate Count}}                                                                                                                                                          & \multicolumn{5}{c|}{\textbf{Circuit Depth}}                                                                                                                                                              \\ \cline{3-17} 
\multicolumn{1}{|c|}{}                                   & \multicolumn{1}{c||}{}                                & \multicolumn{1}{l|}{\textbf{\textsf{JW}}} & \multicolumn{1}{l|}{\textbf{\textsf{BK}}} & \multicolumn{1}{l|}{\textbf{\textsf{BTT}}} & \multicolumn{1}{l|}{\textbf{\textsf{FH}}} & \textbf{\textsf{\name}} & \multicolumn{1}{l|}{\textbf{\textsf{JW}}} & \multicolumn{1}{l|}{\textbf{\textsf{BK}}} & \multicolumn{1}{l|}{\textbf{\textsf{BTT}}} & \multicolumn{1}{l|}{\textbf{\textsf{FH}}} & \textbf{\textsf{\name}} & \multicolumn{1}{l|}{\textbf{\textsf{JW}}} & \multicolumn{1}{l|}{\textbf{\textsf{BK}}} & \multicolumn{1}{l|}{\textbf{\textsf{BTT}}} & \multicolumn{1}{l|}{\textbf{\textsf{FH}}} & \textbf{\textsf{\name}} \\ \hline
$H_2$ \revise{sto3g}                                                    & 4                                                    & \multicolumn{1}{l|}{\textbf{32}}          & \multicolumn{1}{l|}{34}                   & \multicolumn{1}{l|}{36}                    & \multicolumn{1}{l|}{\textbf{32}}          & \textbf{32}             & \multicolumn{1}{l|}{\textbf{21}}          & \multicolumn{1}{l|}{25}                   & \multicolumn{1}{l|}{32}                    & \multicolumn{1}{l|}{22}                   & \textbf{21}             & \multicolumn{1}{l|}{34}                   & \multicolumn{1}{l|}{25}                   & \multicolumn{1}{l|}{53}                    & \multicolumn{1}{l|}{\textbf{33}}          & 34                      \\ \hline
$LiH$ \revise{sto3g frz}                                           & 6                                                    & \multicolumn{1}{l|}{192}                  & \multicolumn{1}{l|}{221}                  & \multicolumn{1}{l|}{225}                   & \multicolumn{1}{l|}{193*}                 & \textbf{188}            & \multicolumn{1}{l|}{\textbf{134}}                  & \multicolumn{1}{l|}{211}                  & \multicolumn{1}{l|}{225}                   & \multicolumn{1}{l|}{189*}                 & 147            & \multicolumn{1}{l|}{\textbf{191}}                  & \multicolumn{1}{l|}{283}                  & \multicolumn{1}{l|}{305}                   & \multicolumn{1}{l|}{254*}                 & 198            \\ \hline
$LiH$ \revise{sto3g}                                                  & 12                                                   & \multicolumn{1}{l|}{3660}                 & \multicolumn{1}{l|}{3248}                 & \multicolumn{1}{l|}{3536}                  & \multicolumn{1}{l|}{3842*}                & \textbf{2926}           & \multicolumn{1}{l|}{2377}                 & \multicolumn{1}{l|}{2373}                 & \multicolumn{1}{l|}{2298}                  & \multicolumn{1}{l|}{2985*}                & \textbf{1642}           & \multicolumn{1}{l|}{3174}                 & \multicolumn{1}{l|}{3249}                 & \multicolumn{1}{l|}{3306}                  & \multicolumn{1}{l|}{3987*}                & \textbf{2402}           \\ \hline
$H_2O$ \revise{sto3g}                                                  & 14                                                   & \multicolumn{1}{l|}{6332}                 & \multicolumn{1}{l|}{6567}                 & \multicolumn{1}{l|}{6658}                  & \multicolumn{1}{l|}{--}                   & \textbf{5545}           & \multicolumn{1}{l|}{4620}                 & \multicolumn{1}{l|}{5064}                 & \multicolumn{1}{l|}{4413}                  & \multicolumn{1}{l|}{--}                   & \textbf{3083}           & \multicolumn{1}{l|}{5755}                 & \multicolumn{1}{l|}{6501}                 & \multicolumn{1}{l|}{6234}                  & \multicolumn{1}{l|}{--}                   & \textbf{4344}           \\ \hline
$CH_4$ \revise{sto3g}                                                   & 18                                                   & \multicolumn{1}{l|}{42476}                & \multicolumn{1}{l|}{42646}                & \multicolumn{1}{l|}{41530}                 & \multicolumn{1}{l|}{--}                   & \textbf{36983}          & \multicolumn{1}{l|}{22798}                & \multicolumn{1}{l|}{19051}                & \multicolumn{1}{l|}{18645}                 & \multicolumn{1}{l|}{--}                   & \textbf{16304}          & \multicolumn{1}{l|}{32951}                & \multicolumn{1}{l|}{28704}                & \multicolumn{1}{l|}{28326}                 & \multicolumn{1}{l|}{--}                   & \textbf{25745}          \\ \hline
$O_2$ \revise{sto3g}                                                    & 20                                                   & \multicolumn{1}{l|}{16904}                & \multicolumn{1}{l|}{16828}                & \multicolumn{1}{l|}{15364}                 & \multicolumn{1}{l|}{--}                   & \textbf{13076}          & \multicolumn{1}{l|}{11653}                & \multicolumn{1}{l|}{12126}                & \multicolumn{1}{l|}{10382}                 & \multicolumn{1}{l|}{--}                   & \textbf{8912}           & \multicolumn{1}{l|}{14532}                & \multicolumn{1}{l|}{15426}                & \multicolumn{1}{l|}{13914}                 & \multicolumn{1}{l|}{--}                   & \textbf{11331}          \\ \hline
$NaF$ \revise{sto3g}                                                   & 28                                                   & \multicolumn{1}{l|}{247264}               & \multicolumn{1}{l|}{218688}               & \multicolumn{1}{l|}{207554}                & \multicolumn{1}{l|}{--}                   & \textbf{192064}         & \multicolumn{1}{l|}{102689}               & \multicolumn{1}{l|}{93376}                & \multicolumn{1}{l|}{89260}                 & \multicolumn{1}{l|}{--}                   & \textbf{69243}          & \multicolumn{1}{l|}{141317}               & \multicolumn{1}{l|}{141577}               & \multicolumn{1}{l|}{133931}                & \multicolumn{1}{l|}{--}                   & \textbf{110898}         \\ \hline
$CO_2$ \revise{sto3g}                                                  & 30                                                   & \multicolumn{1}{l|}{173324}               & \multicolumn{1}{l|}{144112}               & \multicolumn{1}{l|}{138756}                & \multicolumn{1}{l|}{--}                   & \textbf{133208}         & \multicolumn{1}{l|}{85711}                & \multicolumn{1}{l|}{70769}                & \multicolumn{1}{l|}{62857}                 & \multicolumn{1}{l|}{--}                   & \textbf{58208}          & \multicolumn{1}{l|}{110321}               & \multicolumn{1}{l|}{98726}                & \multicolumn{1}{l|}{92334}                 & \multicolumn{1}{l|}{--}                   & \textbf{87114}          \\ \hline
\end{tabular}}
\end{table*}

\section{Evaluation}

In this section, we evaluate the proposed Hamiltonian-adaptive Fermion-to-qubit mapping compilation framework \textsf{HATT} against existing Fermion-to-qubit mappings. We also test the scalability and performance of \name{} and its optimization.


\subsection{Benchmark Physics Models}

We select the following physical models as our benchmark Fermionic Hamiltonians. They come from various application domains of quantum simulation and have different Fermionic mode coupling structures. 
\begin{enumerate}
    \item \textbf{Electronic structure model} from quantum chmistry~\cite{mcquarrie2008quantum}. Hamiltonian describes the electrons in a molecule:
    \begin{equation*}
        \mathcal{H}_\text{e}=\sum_{p,q}h_{pq}a^\dagger_pa_q+\frac{1}{2}\sum_{p,q,r,s}h_{pqrs}a^\dagger_pa^\dagger_qa_ra_s
    \end{equation*}
    Geometric data of molecules are from PubChem~\cite{kim2023pubchem} to determine the coefficients using PySCF~\cite{sun2018pyscf}.
    \item \textbf{Fermi-Hubbard model} in condensed-matter physics~\cite{altland2006condensed}. Hamiltonian describes a position lattice model of Fermions:
    \begin{equation*}
        \mathcal{H}_\text{fh}=\sum_{i,j}\sum_{\sigma=\uparrow,\downarrow}t_{ij}a^\dagger_{i,\sigma}a_{j,\sigma}+U\sum_ia^\dagger_{i,\uparrow}a_{i,\uparrow}a^\dagger_{i,\downarrow}a_{i,\downarrow}
    \end{equation*}
    \item \textbf{Collective neutrino oscillations} from astroparticle physics~\cite{barger2012physics,Patwardhan:2020,Cirigliano:2024}. The Hamiltonian is formulated on a 1D momentum lattice: 
    \begin{equation*}
    \begin{split}
    H_{\nu}=&  \sum_{i=1}^{N}{\sum_{a}^3{\sqrt{p_{i}^2 + m_a^2}} \had_{a,i} \ha_{a,i}}  \\
    +&  \sum_{i_1,i_2,i_3}^{N}  \sum_{a,b}^3 C_{i_1,i_2,i_3}
\had_{a,i_1} 
    \ha_{a,i_3}\had_{b,i_2} \ha_{b,4}
    \end{split}
    \end{equation*}
     where $p_{*,*}$ and $m_a$ are the momentum and static mass of neutrino and $C_{i_1,i_2,i_3}=\mu \left[\hat{p}_{i_2,x} - \hat{p}_{i_1,x}\right]\left[\hat{p}_{4,x} - \hat{p}_{i_3,x}\right]$ ($\mu$ is the two-body coupling strength). 

\end{enumerate}


\subsection{Experiment Setup}


\subsubsection{Implementation}

We implemented our Hamiltonian-adaptive Ternary Tree method (\textsf{\name}) as described in Section~\ref{sec:tree-construction} and~\ref{sec:optimized-algorithm} using Python. We leveraged some Pauli operator processing modules from Qiskit~\cite{qiskit} and Qiskit Nature~\cite{qiskit-nature}.

\subsubsection{Baseline Fermion-to-Qubit Mappings}

We compared against a) the Jordan-Wigner transformation (\textsf{JW})~\cite{jordan1928uber}, and b) the Bravyi-Kitaev transformation (\textsf{BK})~\cite{bravyi2002fermionic} in Qiskit Nature~\cite{qiskit-nature}, c) the balanced ternary tree mapping~\cite{miller2023bonsai} (\textsf{BTT}), and Fermihedral~\cite{liu2024fermihedral} (\textsf{FH}) that gives optimal Hamiltonian Pauli weight at small scales using a SAT solver.


\subsubsection{Compilation}

The time evolution operator is compiled and optimized with the quantum simulation kernel compiler Paulihedral~\cite{li2022paulihedral}, \revise{Rustiq~\cite{debrugière2024fastershortersynthesishamiltonian} or Tetris~\cite{jin2024tetriscompilationframeworkvqa}} followed by Qiskit L3 optimization. We chose $\{CNOT,U3\}$ as the basis gates for noisy simulation. \revise{For Tetris~\cite{jin2024tetriscompilationframeworkvqa}, we tested against the IBM \textit{Manhattan}, \textit{Sycamore}, and \textit{Montreal} superconducting architectures.}


\subsubsection{Noisy Simulation}

We use the Qiskit Aer~\cite{qiskit} to simulate circuits with depolarizing noise on single- and two-qubit gates. 


\subsubsection{Real-System Study}

We also executed the compiled circuit on the IonQ Forte 1 quantum computer. It has 36 ion trap qubits with all-to-all connectivity, $99.98\%$ single-qubit gate fidelity, $98.99\%$ double-qubit gate fidelity, and $99.02\%$ readout fidelity. Due to current hardware limitations, only the $H_2$ molecule simulation is executed.


\subsubsection{Metrics}

We use the following metrics: 1) Pauli weight of the mapped qubit Hamiltonian, 2) $CNOT$ gate count and depth of the compiled circuit, and 3) the simulated system energy in noisy simulation and real-system study. 


\subsection{Pauli Weight and Circuit Metrics}
We evaluate the Pauli weight of the qubit Hamiltonians generated by different Fermion-to-qubit mapping methods and the circuits to simulate these Hamiltonians.

\subsubsection{Electronic Structure Model}

Table~\ref{tab:electronic-strcuture} shows the Pauli weight and circuit metrics of different molecules. For the smallest case where Fermihedral (\textsf{FH}) can find the optimal Pauli weight, \textsf{\name} achieves similar results. For larger cases where \textsf{FH} can only provide approximate solutions or even fail to solve, \textsf{\name} consistently shows the best results in all metrics except for the 6-mode $LiH$ sto3g frz where Jordan-Wigner transformation (\textsf{JW}) is slightly better. Compared with (\textsf{JW}), \textsf{\name}, on average, reduces Pauli weight by $14.77\%$, number of $CNOT$ gates by $25.84\%$, and circuit depth by $19.33\%$. Compared with the Bravyi-Kitaev transformation (\textsf{BK}), \textsf{\name} reduces $13.83\%$ Pauli weight, $24.35\%$ $CNOT$ count and $18.91\%$ circuit depth. Compared gainst the balanced ternary tree (\textsf{BTT}), \textsf{\name} reduces $11.77\%$ Pauli weight, $21.37\%$ $CNOT$ count and $20.58\%$ circuit depth. \revise{Compared with \textsf{JW} on more sophisticated workflows, like Rustiq~\cite{debrugière2024fastershortersynthesishamiltonian} and Tetris~\cite{jin2024tetriscompilationframeworkvqa}, \textsf{\name{}} still demonstrates better performance for almost all cases (results shown in Table~\ref{tab:exp-extra-rustiq} and~\ref{tab:exp-extra-tetris}). Compared with \textsf{JW}+Rustiq, \textsf{\name{}}+Rustiq further reduces up to 18.2\% $CNOT$ count, 21.83\% $U3$ count, and 13.5\% circuit depth. For compilation onto superconducting architectures (\textit{Manhattan}, \textit{Sycamore} and \textit{Montreal}), \textsf{\name{}}+Tetris can outperform \textsf{JW}+Tetris with reduction on $CNOT$ count, $U3$ count, and circuit depth by up to $17.11\%$, $22.00\%$ and $19.52\%$, respectively. Note that both Rustiq and Tetris are developed using \textsf{JW}. It is possible to further optimize the workflow for the Pauli strings patterns generated by \textsf{\name{}}.}

\begin{table*}[t]
\centering
\caption{Evaluation result of Fermi-Hubbard Model. `--' indicates the case is too large to solve by Fermihedral (\textsf{FH}).}
\label{tab:fermi-hubbard}
\resizebox{\linewidth}{!}{%
\begin{tabular}{|l|l||lllll||lllll||lllll|}
\hline
\multicolumn{1}{|c|}{\multirow{2}{*}{\textbf{Geometry}}} & \multicolumn{1}{c||}{\multirow{2}{*}{\textbf{Modes}}} & \multicolumn{5}{c||}{\textbf{Pauli Weight}}                                                                                                                                                              & \multicolumn{5}{c||}{\textbf{$CNOT$ Gate Count}}                                                                                                                                                         & \multicolumn{5}{c|}{\textbf{Circuit Depth}}                                                                                                                                                             \\ \cline{3-17} 
\multicolumn{1}{|c|}{}                               & \multicolumn{1}{c||}{}                                & \multicolumn{1}{l|}{\textbf{\textsf{JW}}} & \multicolumn{1}{l|}{\textbf{\textsf{BK}}} & \multicolumn{1}{l|}{\textbf{\textsf{BTT}}} & \multicolumn{1}{l|}{\textbf{\textsf{FH}}} & \textbf{\textsf{HATT}} & \multicolumn{1}{l|}{\textbf{\textsf{JW}}} & \multicolumn{1}{l|}{\textbf{\textsf{BK}}} & \multicolumn{1}{l|}{\textbf{\textsf{BTT}}} & \multicolumn{1}{l|}{\textbf{\textsf{FH}}} & \textbf{\textsf{HATT}} & \multicolumn{1}{l|}{\textbf{\textsf{JW}}} & \multicolumn{1}{l|}{\textbf{\textsf{BK}}} & \multicolumn{1}{l|}{\textbf{\textsf{BTT}}} & \multicolumn{1}{l|}{\textbf{\textsf{FH}}} & \textbf{\textsf{HATT}} \\ \hline
$2\times2$                                                  & 8                                                    & \multicolumn{1}{l|}{80}                   & \multicolumn{1}{l|}{80}                   & \multicolumn{1}{l|}{86}                    & \multicolumn{1}{l|}{\textbf{56}}          & 76                     & \multicolumn{1}{l|}{51}                   & \multicolumn{1}{l|}{71}                   & \multicolumn{1}{l|}{77}                    & \multicolumn{1}{l|}{\textbf{37}}          & 62                     & \multicolumn{1}{l|}{60}                   & \multicolumn{1}{l|}{99}                   & \multicolumn{1}{l|}{108}                   & \multicolumn{1}{l|}{\textbf{36}}          & 73                     \\ \hline
$2\times3$                                                  & 12                                                   & \multicolumn{1}{l|}{212}                  & \multicolumn{1}{l|}{200}                  & \multicolumn{1}{l|}{199}                   & \multicolumn{1}{l|}{\textbf{161}}         & 187                    & \multicolumn{1}{l|}{159}                  & \multicolumn{1}{l|}{172}                  & \multicolumn{1}{l|}{161}                   & \multicolumn{1}{l|}{\textbf{123}}         & 146                    & \multicolumn{1}{l|}{161}                  & \multicolumn{1}{l|}{223}                  & \multicolumn{1}{l|}{229}                   & \multicolumn{1}{l|}{\textbf{160}}         & 196                    \\ \hline
$2\times4$                                                  & 16                                                   & \multicolumn{1}{l|}{304}                  & \multicolumn{1}{l|}{263}                  & \multicolumn{1}{l|}{260}                   & \multicolumn{1}{l|}{\textbf{230}}         & 256                    & \multicolumn{1}{l|}{228}                  & \multicolumn{1}{l|}{\textbf{183}}         & \multicolumn{1}{l|}{208}                   & \multicolumn{1}{l|}{195}                  & 189                    & \multicolumn{1}{l|}{\textbf{225}}         & \multicolumn{1}{l|}{239}                  & \multicolumn{1}{l|}{256}                   & \multicolumn{1}{l|}{230}                  & 249                    \\ \hline
$3\times3$                                                  & 18                                                   & \multicolumn{1}{l|}{492}                  & \multicolumn{1}{l|}{428}                  & \multicolumn{1}{l|}{408}                   & \multicolumn{1}{l|}{\textbf{352}}         & 410                    & \multicolumn{1}{l|}{378}                  & \multicolumn{1}{l|}{296}                  & \multicolumn{1}{l|}{317}                   & \multicolumn{1}{l|}{266}                  & \textbf{260}           & \multicolumn{1}{l|}{328}                  & \multicolumn{1}{l|}{391}                  & \multicolumn{1}{l|}{427}                   & \multicolumn{1}{l|}{\textbf{270}}         & 356                    \\ \hline
$2\times5$                                                  & 20                                                   & \multicolumn{1}{l|}{396}                  & \multicolumn{1}{l|}{348}                  & \multicolumn{1}{l|}{356}                   & \multicolumn{1}{l|}{--}                   & \textbf{330}           & \multicolumn{1}{l|}{287}                  & \multicolumn{1}{l|}{270}                  & \multicolumn{1}{l|}{266}                   & \multicolumn{1}{l|}{--}                   & \textbf{224}           & \multicolumn{1}{l|}{305}                  & \multicolumn{1}{l|}{\textbf{275}}         & \multicolumn{1}{l|}{320}                   & \multicolumn{1}{l|}{--}                   & 283                    \\ \hline
$3\times4$                                                  & 24                                                   & \multicolumn{1}{l|}{704}                  & \multicolumn{1}{l|}{620}                  & \multicolumn{1}{l|}{580}                   & \multicolumn{1}{l|}{--}                   & \textbf{524}           & \multicolumn{1}{l|}{528}                  & \multicolumn{1}{l|}{460}                  & \multicolumn{1}{l|}{433}                   & \multicolumn{1}{l|}{--}                   & \textbf{364}           & \multicolumn{1}{l|}{462}                  & \multicolumn{1}{l|}{529}                  & \multicolumn{1}{l|}{496}                   & \multicolumn{1}{l|}{--}                   & \textbf{401}           \\ \hline
$2\times7$                                                  & 28                                                   & \multicolumn{1}{l|}{580}                  & \multicolumn{1}{l|}{493}                  & \multicolumn{1}{l|}{502}                   & \multicolumn{1}{l|}{--}                   & \textbf{473}           & \multicolumn{1}{l|}{405}                  & \multicolumn{1}{l|}{380}                  & \multicolumn{1}{l|}{373}                   & \multicolumn{1}{l|}{--}                   & \textbf{320}           & \multicolumn{1}{l|}{465}                  & \multicolumn{1}{l|}{374}                  & \multicolumn{1}{l|}{399}                   & \multicolumn{1}{l|}{--}                   & \textbf{333}           \\ \hline
$3\times5$                                                  & 30                                                   & \multicolumn{1}{l|}{916}                  & \multicolumn{1}{l|}{756}                  & \multicolumn{1}{l|}{\textbf{706}}          & \multicolumn{1}{l|}{--}                   & \textbf{706}           & \multicolumn{1}{l|}{661}                  & \multicolumn{1}{l|}{523}                  & \multicolumn{1}{l|}{563}                   & \multicolumn{1}{l|}{--}                   & \textbf{490}           & \multicolumn{1}{l|}{621}                  & \multicolumn{1}{l|}{\textbf{550}}         & \multicolumn{1}{l|}{629}                   & \multicolumn{1}{l|}{--}                   & 582                    \\ \hline
$4\times4$                                                  & 32                                                   & \multicolumn{1}{l|}{1152}                 & \multicolumn{1}{l|}{790}                  & \multicolumn{1}{l|}{784}                   & \multicolumn{1}{l|}{--}                   & \textbf{760}           & \multicolumn{1}{l|}{842}                  & \multicolumn{1}{l|}{531}                  & \multicolumn{1}{l|}{651}                   & \multicolumn{1}{l|}{--}                   & \textbf{483}           & \multicolumn{1}{l|}{712}                  & \multicolumn{1}{l|}{562}                  & \multicolumn{1}{l|}{749}                   & \multicolumn{1}{l|}{--}                   & \textbf{553}           \\ \hline
$3\times6$                                                  & 36                                                   & \multicolumn{1}{l|}{1128}                 & \multicolumn{1}{l|}{932}                  & \multicolumn{1}{l|}{876}                   & \multicolumn{1}{l|}{--}                   & \textbf{806}           & \multicolumn{1}{l|}{794}                  & \multicolumn{1}{l|}{662}                  & \multicolumn{1}{l|}{668}                   & \multicolumn{1}{l|}{--}                   & \textbf{544}           & \multicolumn{1}{l|}{780}                  & \multicolumn{1}{l|}{684}                  & \multicolumn{1}{l|}{694}                   & \multicolumn{1}{l|}{--}                   & \textbf{487}           \\ \hline
$4\times5$                                                  & 40                                                   & \multicolumn{1}{l|}{1504}                 & \multicolumn{1}{l|}{1030}                 & \multicolumn{1}{l|}{\textbf{986}}          & \multicolumn{1}{l|}{--}                   & \textbf{986}           & \multicolumn{1}{l|}{1055}                 & \multicolumn{1}{l|}{665}                  & \multicolumn{1}{l|}{782}                   & \multicolumn{1}{l|}{--}                   & \textbf{601}           & \multicolumn{1}{l|}{941}                  & \multicolumn{1}{l|}{\textbf{561}}         & \multicolumn{1}{l|}{795}                   & \multicolumn{1}{l|}{--}                   & 618                    \\ \hline
\end{tabular}}
\end{table*}

\begin{table*}[t]
\vspace{-5pt}
\centering
\caption{Evaluation result of Collective Neutrino Oscillation. Fermihedral (FH) is not included since all the benchmarks in this application are too large for Fermihedral.}
\label{tab:neutrino-oscillation}
\resizebox{\linewidth}{!}{%
\begin{tabular}{|l|l||llll||llll||llll|}
\hline
\multicolumn{1}{|c|}{\multirow{2}{*}{\textbf{Case}}} & \multicolumn{1}{c||}{\multirow{2}{*}{\textbf{Modes}}} & \multicolumn{4}{c||}{\textbf{Pauli Weight}}                                                                                                                  & \multicolumn{4}{c||}{\textbf{$CNOT$ Gate Count}}                                                                                                             & \multicolumn{4}{c|}{\textbf{Circuit Depth}}                                                                                                                 \\ \cline{3-14} 
\multicolumn{1}{|c|}{}                      & \multicolumn{1}{c||}{}                       & \multicolumn{1}{l|}{\textbf{\textsf{JW}}} & \multicolumn{1}{l|}{\textbf{\textsf{BK}}} & \multicolumn{1}{l|}{\textbf{\textsf{BTT}}} & \textbf{\textsf{HATT}} & \multicolumn{1}{l|}{\textbf{\textsf{JW}}} & \multicolumn{1}{l|}{\textbf{\textsf{BK}}} & \multicolumn{1}{l|}{\textbf{\textsf{BTT}}} & \textbf{\textsf{HATT}} & \multicolumn{1}{l|}{\textbf{\textsf{JW}}} & \multicolumn{1}{l|}{\textbf{\textsf{BK}}} & \multicolumn{1}{l|}{\textbf{\textsf{BTT}}} & \textbf{\textsf{HATT}} \\ \hline
$3\times2 F$                                & 12                                          & \multicolumn{1}{l|}{1424}                 & \multicolumn{1}{l|}{1568}                 & \multicolumn{1}{l|}{1556}                  & \textbf{1290}          & \multicolumn{1}{l|}{\textbf{776}}         & \multicolumn{1}{l|}{986}                  & \multicolumn{1}{l|}{1092}                  & 850                    & \multicolumn{1}{l|}{\textbf{1137}}        & \multicolumn{1}{l|}{1421}                 & \multicolumn{1}{l|}{1554}                  & 1229                   \\ \hline
$4\times2 F$                                & 16                                          & \multicolumn{1}{l|}{4048}                 & \multicolumn{1}{l|}{4011}                 & \multicolumn{1}{l|}{4244}                  & \textbf{3720}          & \multicolumn{1}{l|}{\textbf{2115}}        & \multicolumn{1}{l|}{2742}                 & \multicolumn{1}{l|}{2657}                  & 2203                   & \multicolumn{1}{l|}{\textbf{3003}}        & \multicolumn{1}{l|}{3763}                 & \multicolumn{1}{l|}{3788}                  & 3110                   \\ \hline
$3\times3 F$                                & 18                                          & \multicolumn{1}{l|}{5550}                 & \multicolumn{1}{l|}{5770}                 & \multicolumn{1}{l|}{5548}                  & \textbf{5153}          & \multicolumn{1}{l|}{2912}                 & \multicolumn{1}{l|}{3667}                 & \multicolumn{1}{l|}{3391}                  & \textbf{2703}          & \multicolumn{1}{l|}{\textbf{3927}}        & \multicolumn{1}{l|}{4911}                 & \multicolumn{1}{l|}{4801}                  & 3949                   \\ \hline
$5\times2 F$                                & 20                                          & \multicolumn{1}{l|}{9240}                 & \multicolumn{1}{l|}{9800}                 & \multicolumn{1}{l|}{9016}                  & \textbf{7852}          & \multicolumn{1}{l|}{4630}                 & \multicolumn{1}{l|}{5285}                 & \multicolumn{1}{l|}{5685}                  & \textbf{4487}          & \multicolumn{1}{l|}{\textbf{6261}}        & \multicolumn{1}{l|}{7476}                 & \multicolumn{1}{l|}{8021}                  & 6414                   \\ \hline
$4\times3 F$                                & 24                                          & \multicolumn{1}{l|}{16216}                & \multicolumn{1}{l|}{16462}                & \multicolumn{1}{l|}{14806}                 & \textbf{14267}         & \multicolumn{1}{l|}{7996}                 & \multicolumn{1}{l|}{8952}                 & \multicolumn{1}{l|}{8243}                  & \textbf{7141}          & \multicolumn{1}{l|}{10530}                & \multicolumn{1}{l|}{12098}                & \multicolumn{1}{l|}{11786}                 & \textbf{10440}         \\ \hline
$6\times2 F$                                & 24                                          & \multicolumn{1}{l|}{18280}                & \multicolumn{1}{l|}{18594}                & \multicolumn{1}{l|}{16992}                 & \textbf{15047}         & \multicolumn{1}{l|}{8868}                 & \multicolumn{1}{l|}{9168}                 & \multicolumn{1}{l|}{9612}                  & \textbf{8382}          & \multicolumn{1}{l|}{\textbf{11571}}       & \multicolumn{1}{l|}{13338}                & \multicolumn{1}{l|}{13693}                 & 11986                  \\ \hline
$7\times2 F$                                & 28                                          & \multicolumn{1}{l|}{32704}                & \multicolumn{1}{l|}{31088}                & \multicolumn{1}{l|}{28876}                 & \textbf{25074}         & \multicolumn{1}{l|}{15440}                & \multicolumn{1}{l|}{14733}                & \multicolumn{1}{l|}{15358}                 & \textbf{13322}         & \multicolumn{1}{l|}{\textbf{19281}}       & \multicolumn{1}{l|}{21278}                & \multicolumn{1}{l|}{22148}                 & 19400                  \\ \hline
$5\times3 F$                                & 30                                          & \multicolumn{1}{l|}{37690}                & \multicolumn{1}{l|}{33776}                & \multicolumn{1}{l|}{32154}                 & \textbf{31418}         & \multicolumn{1}{l|}{17872}                & \multicolumn{1}{l|}{17460}                & \multicolumn{1}{l|}{16957}                 & \textbf{15204}         & \multicolumn{1}{l|}{21958}                & \multicolumn{1}{l|}{23708}                & \multicolumn{1}{l|}{23872}                 & \textbf{21697}         \\ \hline
$6\times3 F$                                & 36                                          & \multicolumn{1}{l|}{75540}                & \multicolumn{1}{l|}{66262}                & \multicolumn{1}{l|}{60576}                 & \textbf{58229}         & \multicolumn{1}{l|}{34697}                & \multicolumn{1}{l|}{30193}                & \multicolumn{1}{l|}{29361}                 & \textbf{26298}         & \multicolumn{1}{l|}{41198}                & \multicolumn{1}{l|}{41702}                & \multicolumn{1}{l|}{41995}                 & \textbf{38502}         \\ \hline
$7\times3 F$                                & 42                                          & \multicolumn{1}{l|}{136486}               & \multicolumn{1}{l|}{114833}               & \multicolumn{1}{l|}{101717}                & \textbf{99334}         & \multicolumn{1}{l|}{60414}                & \multicolumn{1}{l|}{48846}                & \multicolumn{1}{l|}{48155}                 & \textbf{45045}         & \multicolumn{1}{l|}{69117}                & \multicolumn{1}{l|}{67548}                & \multicolumn{1}{l|}{67600}                 & \textbf{64686}         \\ \hline
\end{tabular}}
\vspace{-10pt}
\end{table*}

\subsubsection{Fermi-Hubbard Model}

Table~\ref{tab:fermi-hubbard} shows the Pauli weight and two metrics of the compiled circuits for the Fermi-Hubbard model benchmarks. It can be observed that although \textsf{FH} achieved the minimal Pauli weight for up to $18$ modes, it cannot solve more significant cases. \textsf{\name} achieves a lower Pauli weight consistently and a lower circuit complexity most of the time against \textsf{JW}, \textsf{BK}, and \textsf{BTT} as our method is tailored according to a specific Hamiltonian and does capture the optimization opportunities within.

On Fermi-Hubbard models, compared to \textsf{JW}, \textsf{\name} on average reduces Pauli weight by $20.90\%$, number of $CNOT$ gates by $22.90\%$, and circuit depth by $7.88\%$. Compared to \textsf{BK}, \textsf{\name} reduces $6.48\%$ Pauli weight, $12.11\%$ $CNOT$ count and $8.16\%$ circuit depth. Against \textsf{BTT}, \textsf{\name} reduces $4.77\%$ Pauli weight, $16.58\%$ $CNOT$ count and $18.11\%$ circuit depth correspondingly. On small scales ($8\sim18$ modes), \textsf{\name} shows the results closest to the optimal solution by \textsf{FH}.


\subsubsection{Collective Neutrino Oscillation}

Table~\ref{tab:neutrino-oscillation} shows the Pauli weight and metrics of compiled circuits of different neutrino oscillation test cases. Fermihedral (\textsf{FH}) is not evaluated since all the cases are too large for \textsf{FH}. Among all the cases, \textsf{\name} always achieves the lower Pauli weight.
Although \textsf{\name} shows slightly higher circuit overhead compared with \textsf{JW}, the trend shows that \textsf{\name} has more advantage as the problem size increases.
Compared to \textsf{JW}, \textsf{\name} on average reduces Pauli weight by $15.65\%$, number of $CNOT$ gates by $4.01\%$. Compared to \textsf{BK}, \textsf{\name} reduces $14.58\%$ Pauli weight, $17.69\%$ $CNOT$ count and $14.92\%$ circuit depth. Against \textsf{BTT}, \textsf{\name} reduces $11.95\%$ Pauli weight, $17.30\%$ $CNOT$ count and $17.31\%$ circuit depth correspondingly. 

\begin{table}[ht]
  \centering
  \caption{\revise{Evaluation result of Electronic Structure Model with \textbf{Tetris}~\cite{jin2024tetriscompilationframeworkvqa} on Manhattan, Sycamore and Montreal}}
  \label{tab:exp-extra-tetris}
  \resizebox{\linewidth}{!}{
  \begin{tabular}{|l||l|l||l|l||l|l|}
    \hline
    \multirow{2}{*}{\textbf{Case}} & \multicolumn{2}{c||}{\textbf{$CNOT$ Gate Count}} & \multicolumn{2}{c||}{\textbf{$U3$ Gate Count}} & \multicolumn{2}{c|}{\textbf{Circuit Depth}} \\\cline{2-7}
     & \sbf{JW} & \sbf{\name{}} & \sbf{JW} & \sbf{\name{}} & \sbf{JW} & \sbf{\name{}} \\\hline\hline
    \multicolumn{7}{|c|}{\large Manhattan} \\\hline
    $H_2$ sto3g & 40 & \textbf{38} & 47 & \textbf{37} & 62 & \textbf{56} \\\hline
    $H_2$ sto3g frz & 42 & \textbf{36} & \textbf{40} & 43 & 61 & \textbf{57} \\\hline
    $H_2$ 631g & 1479 & \textbf{1420} & 1302 & \textbf{1172} & 1744 & \textbf{1659} \\\hline
    $H_2$ 631g frz & 1506 & \textbf{1314} & 1290 & \textbf{1147} & 1769 & \textbf{1540} \\\hline
    $LiH$ sto3g & 7167 & \textbf{6517} & 5029 & \textbf{4616} & 7302 & \textbf{6890} \\\hline
    $LiH$ sto3g frz & 2753 & \textbf{2282} & 2000 & \textbf{1560} & 2915 & \textbf{2346} \\\hline
    $NH$ sto3g frz & 2824 & \textbf{2415} & 1999 & \textbf{1610} & 2881 & \textbf{2563} \\\hline
    $NH$ 631g & 145396 & \textbf{138719} & 80484 & \textbf{72264} & 126398 & \textbf{123177} \\\hline
    $NH$ 631g frz & 85256 & \textbf{80046} & 48451 & \textbf{46037} & 76164 & \textbf{73964} \\\hline
    $BeH_2$ sto3g frz & 11646 & \textbf{11513} & \textbf{8911} & 8980 & 12711 & \textbf{11959} \\\hline
    $O_2$ sto3g & 48660 & \textbf{46483} & 24231 & \textbf{23460} & \textbf{38985} & 39446 \\\hline\hline
    \multicolumn{7}{|c|}{\large Sycamore} \\\hline
    $H_2$ sto3g & 42 & \textbf{38} & 40 & \textbf{37} & 60 & \textbf{56} \\\hline
    $H_2$ sto3g frz & 42 & \textbf{38} & 40 & \textbf{37} & \textbf{56} & \textbf{56} \\\hline
    $H_2$ 631g & 1120 & \textbf{1032} & 1017 & \textbf{965} & 1348 & \textbf{1228} \\\hline
    $H_2$ 631g frz & 1122 & \textbf{1040} & 1044 & \textbf{988} & 1383 & \textbf{1225} \\\hline
    $LiH$ sto3g & 5760 & \textbf{5665} & 4114 & \textbf{3614} & 6031 & \textbf{5941} \\\hline
    $LiH$ sto3g frz & 2098 & \textbf{2001} & 1472 & \textbf{1295} & 2202 & \textbf{2183} \\\hline
    $NH$ sto3g frz & 2214 & \textbf{1976} & 1540 & \textbf{1364} & 2356 & \textbf{2140} \\\hline
    $NH$ 631g & 98096 & \textbf{94642} & 61052 & \textbf{57630} & 89638 & \textbf{86487} \\\hline
    $NH$ 631g frz & 60009 & \textbf{56304} & 38446 & \textbf{33891} & 56151 & \textbf{52170} \\\hline
    $BeH_2$ sto3g frz & 9607 & \textbf{9350} & 7155 & \textbf{6895} & 10427 & \textbf{9805} \\\hline
    $O_2$ sto3g & 31986 & \textbf{29690} & 19296 & \textbf{17390} & 28301 & \textbf{27031} \\\hline\hline
    \multicolumn{7}{|c|}{\large Montreal} \\\hline
    $H_2$ sto3g & 42 & \textbf{36} & \textbf{40} & 43 & 61 & \textbf{57} \\\hline
    $H_2$ sto3g frz & 42 & \textbf{38} & 40 & \textbf{37} & 61 & \textbf{56} \\\hline
    $H_2$ 631g & 1409 & \textbf{1271} & 1324 & \textbf{1140} & 1633 & \textbf{1510} \\\hline
    $H_2$ 631g frz & 1435 & \textbf{1417} & 1300 & \textbf{1160} & 1699 & \textbf{1641} \\\hline
    $LiH$ sto3g & 7057 & \textbf{6465} & 4834 & \textbf{4731} & 7171 & \textbf{6811} \\\hline
    $LiH$ sto3g frz & 2865 & \textbf{2522} & 1949 & \textbf{1628} & 3013 & \textbf{2553} \\\hline
    $NH$ sto3g frz & 2862 & \textbf{2467} & 1939 & \textbf{1524} & 2999 & \textbf{2523} \\\hline
    $NH$ 631g & 146957 & \textbf{144169} & 77300 & \textbf{74026} & 127955 & \textbf{124624} \\\hline
    $NH$ 631g frz & 87792 & \textbf{84268} & 47537 & \textbf{45356} & 77307 & \textbf{74992} \\\hline
    $BeH_2$ sto3g frz & \textbf{11798} & 12350 & 8913 & \textbf{8412} & 12783 & \textbf{12318} \\\hline
    $O_2$ sto3g & 50641 & \textbf{46443} & 24294 & \textbf{22969} & 40857 & \textbf{38649} \\\hline
  \end{tabular}
  }
  \vspace{-10pt}
\end{table}

\begin{table}[ht]
  \centering
  \caption{\revise{Evaluation result of Electronic Structure Model with \textbf{Rustiq}~\cite{debrugière2024fastershortersynthesishamiltonian}}}
  \label{tab:exp-extra-rustiq}
  \resizebox{\linewidth}{!}{
  \begin{tabular}{|l||l|l||l|l||l|l|}
    \hline
    \multirow{2}{*}{\textbf{Case}} & \multicolumn{2}{c||}{\textbf{$CNOT$ Gate Count}} & \multicolumn{2}{c||}{\textbf{$U3$ Gate Count}} & \multicolumn{2}{c|}{\textbf{Circuit Depth}} \\\cline{2-7}
     & \sbf{JW} & \sbf{\name{}} & \sbf{JW} & \sbf{\name{}} & \sbf{JW} & \sbf{\name{}} \\\hline
    $H_2$ sto3g & \textbf{7} & \textbf{7} & 6 & \textbf{2} & \textbf{7} & \textbf{7} \\\hline
    $H_2$ sto3g frz & \textbf{7} & \textbf{7} & 6 & \textbf{2} & \textbf{7} & \textbf{7} \\\hline
    $H_2$ 631g & \textbf{217} & 227 & 180 & \textbf{164} & 164 & \textbf{161} \\\hline
    $H_2$ 631g frz & 217 & \textbf{216} & 180 & \textbf{170} & 164 & \textbf{149} \\\hline
    $LiH$ sto3g & 1348 & \textbf{1230} & 1031 & \textbf{872} & 625 & \textbf{561} \\\hline
    $LiH$ sto3g frz & 423 & \textbf{413} & 318 & \textbf{306} & 252 & \textbf{250} \\\hline
    $LiH$ 631g & 52363 & \textbf{49528} & 42775 & \textbf{39288} & 10638 & \textbf{10243} \\\hline
    $H_2O$ sto3g frz & 1914 & \textbf{1793} & 1580 & \textbf{1481} & 1926 & \textbf{1748} \\\hline
    $H_2O$ 631g & 114393 & \textbf{111815} & 109820 & \textbf{106944} & 96820 & \textbf{95123} \\\hline
    $H_2O$ 631g frz & 107618 & \textbf{104943} & 86674 & \textbf{81105} & 20030 & \textbf{19854} \\\hline
    $NH$ sto3g & 1348 & \textbf{1285} & 1031 & \textbf{906} & 625 & \textbf{588} \\\hline
    $NH$ sto3g frz & 423 & \textbf{346} & 318 & \textbf{308} & 252 & \textbf{225} \\\hline
    $NH$ 631g frz & 22666 & \textbf{18993} & 17726 & \textbf{13856} & 5268 & \textbf{4554} \\\hline
    $BeH_2$ 631g & 168863 & \textbf{162917} & 135486 & \textbf{129317} & 29096 & \textbf{28040} \\\hline
    $BeH_2$ 631g frz & 106886 & \textbf{102840} & 82695 & \textbf{79097} & 20078 & \textbf{19462} \\\hline
    $CH_4$ sto3g & 16641 & \textbf{15741} & 14429 & \textbf{13343} & 15074 & \textbf{14140} \\\hline
    $CH_4$ sto3g frz & 10364 & \textbf{10095} & 7948 & \textbf{7698} & 3344 & \textbf{3332} \\\hline
    $CH_4$ 631g & 852020 & \textbf{796853} & 701829 & \textbf{655846} & 108033 & \textbf{102717} \\\hline
    $O_2$ sto3g & 12012 & \textbf{10592} & 9526 & \textbf{7529} & 2766 & \textbf{2582} \\\hline
    $O_2$ sto3g frz & 3155 & \textbf{3103} & \textbf{2313} & 2374 & 990 & \textbf{944} \\\hline
    $NaF$ sto3g & 204673 & \textbf{203415} & 162845 & \textbf{162837} & 32126 & \textbf{31928} \\\hline
    $NaF$ sto3g frz & 6094 & \textbf{5811} & 5271 & \textbf{5037} & 5759 & \textbf{5472} \\\hline
  \end{tabular}
  }
  \vspace{-10pt}
\end{table}

\subsection{Noisy Simulation and Real-System Study}

We performed the noisy simulation and real-system study of compiled circuits for $H_2$ and $LiH$ sto3g frz cases from the electronic structure model.
`frz' means that the inner layer electrons are fixed.

\begin{figure}[t]
    \centering
    \includegraphics[width=0.93\linewidth]{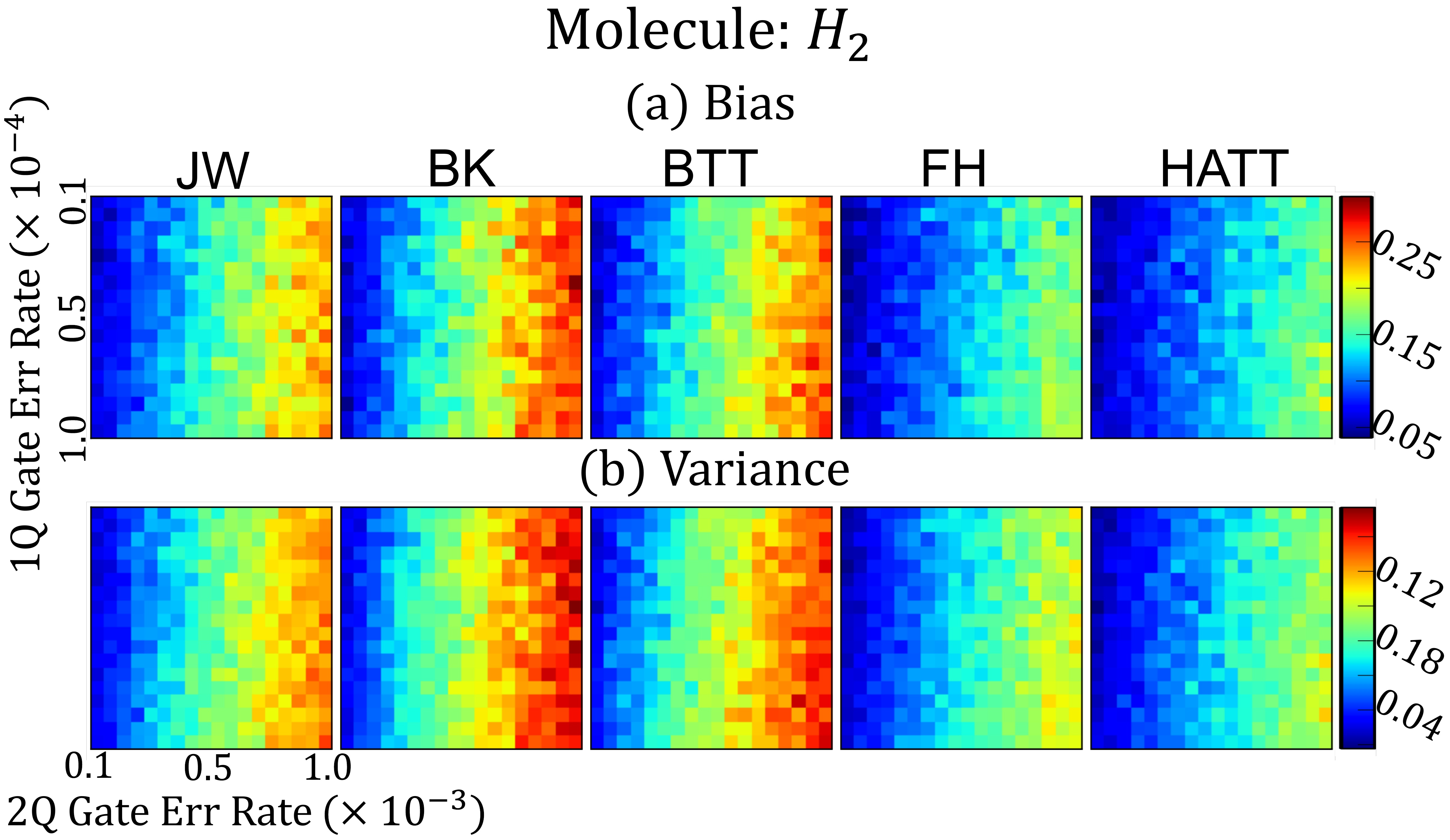}
    \vspace{10pt}
    \centering
    \includegraphics[width=0.93\linewidth]{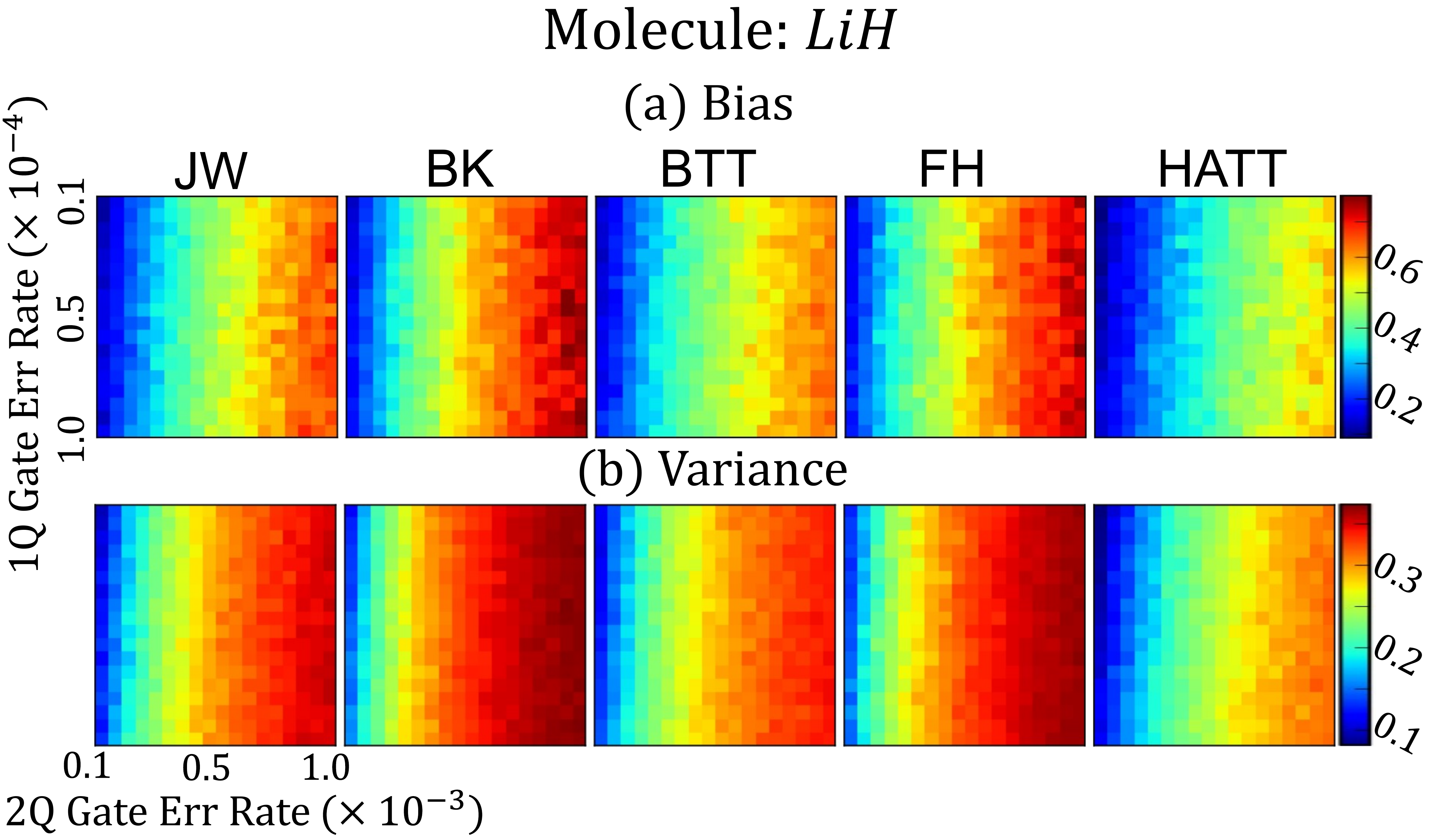}
    \caption{Noisy simulation result of $H_2$ and $LiH$ sto3g frz molecule. Energy is measured for 1000 shots. Bias is calculated against the theoretical results.}
    \vspace{-10pt}
    \label{fig:noisy-simulation}
\end{figure}

\subsubsection{Noisy Simulation}

We simulated the circuit generated by different Fermion-to-qubit mappings under depolarizing errors. The error rate range of single-qubit gates is $10^{-5}\sim10^{-4}$ and $10^{-4}\sim10^{-3}$ for double-qubit gates. We simulate each circuit for 1000 shots, and the final system energy is measured and compared against the theoretical results.

Figure~\ref{fig:noisy-simulation} shows the simulation results of $H_2$ (upper half) and $LiH$ sto3g frz (lower half) molecules. Bias and variance are calculated against the theoretical values based on 1000 shots. It can be observed that \textsf{\name} has the lowest deviation and variance (heatmap closer to blue), which outperforms \textsf{JW}, \textsf{BK}, and \textsf{BTT}, achieving similar results with the optimal \textsf{FH}.

\subsubsection{Real-System Study}

\begin{figure}[t]
    \centering
    \includegraphics[width=0.8\linewidth]{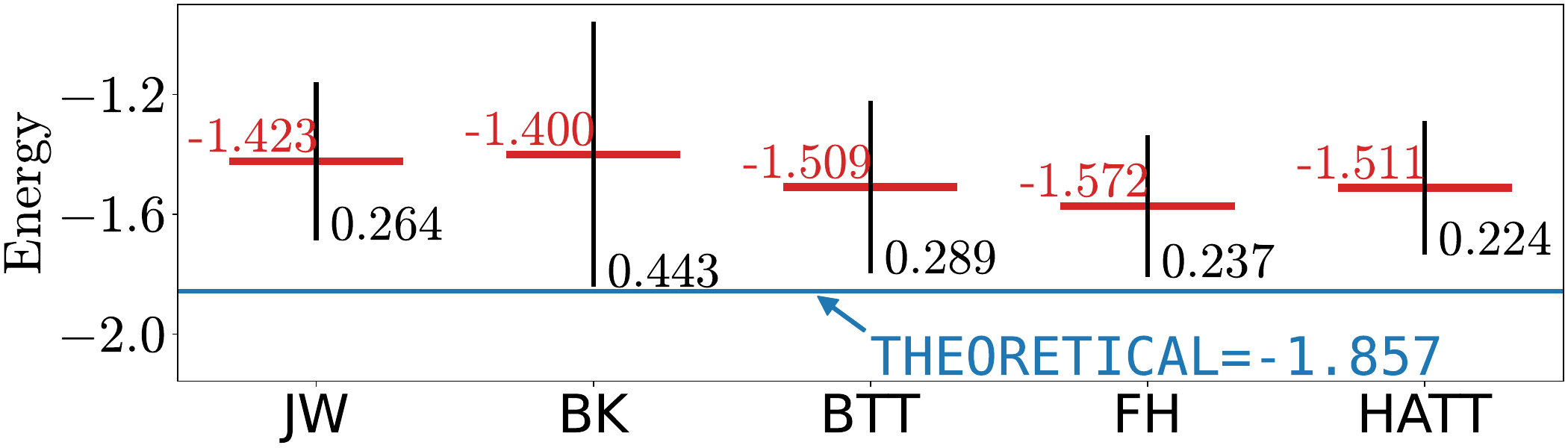}
    \caption{$H_2$ molecule simulation results on IonQ Forte 1 quantum computer}
    \label{fig:result-realsystem}
    \vspace{-10pt}
\end{figure}

Figure~\ref{fig:result-realsystem} shows the results of $H_2$ ground state energy simulation on the IonQ Forte 1 device. The red horizontal lines indicate the average measured energy of 1000 shots using the corresponding mapping. The associated black vertical lines indicate the variance of the measured energy across all shots. The blue line indicates the theoretical result. \textsf{\name} achieved the second closest to theory, with the closest being the small-scale optimal \textsf{FH} solution. \textsf{\name} also has the lowest variance.


\subsection{Execution Time and Scalability}

\begin{figure}[htbp]
    \centering
    \includegraphics[width=0.8\linewidth]{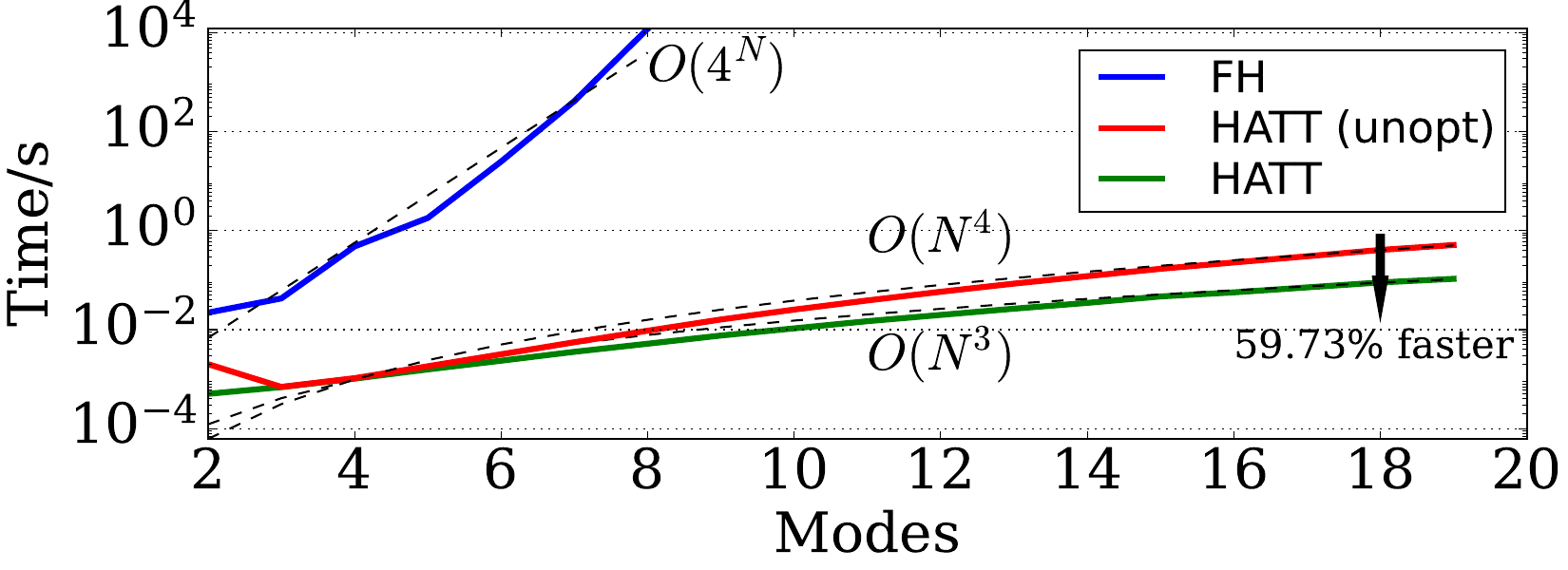}
    \caption{Scalability of Fermihedral (\textsf{FH}), unoptimized HATT (\textsf{HATT (unopt)}) and HATT with optimization (\textsf{HATT}). Black dotted lines (\textsf{- \!\!\!- \!\!\!-}) show the regression of complexity orders.}
    \label{fig:scalability}
    \vspace{-10pt}
\end{figure}

To understand the scalability of the proposed tree construction algorithm and the impact of our performance optimization, we compared our unoptimized tree construction algorithm in Section~\ref{sec:naive-algorithm} (\textsf{HATT (unopt)}), optimized algorithm in Section~\ref{sec:optimized-pairing} (\textsf{HATT}) and the optimal exhaustive search method Fermihedral~\cite{liu2024fermihedral} (\textsf{FH}) on the time consumption to produce a Fermion-to-qubit mapping for a simple Hamiltonian: $\mathcal{H}_\mathcal{F}=\sum_{i=0}^{2N-1} M_i$ at different sizes.
As shown in Figure~\ref{fig:scalability}, both \textsf{HATT (unopt)} and \textsf{HATT} scales to large cases in polynomial time, but \textsf{FH} show its exponential complexity. Also, \textsf{HATT} has a shorter execution time compared with unoptimized \textsf{HATT (unopt)} as our optimization reduces the algorithm's complexity from $O(N^4)$ to $O(N^3)$.


\subsection{\revise{Performance Cost of Optimized \name{}}}

\revise{We also compared the Pauli weight in Table~\hyperref[tab:performance-cost]{VI} of \textsf{HATT (unopt)} and \textsf{HATT} to show how optimization could affect performance. The result shows that \textsf{HATT} and \textsf{HATT (unopt)} generate Hamiltonians with $\sim0.43\%$ difference in Pauli weight on average. Thus, the optimization does not come with a significant performance trade-off.}

\begin{table}[ht]
\label{tab:performance-cost}
\caption{\revise{Comparison of Pauli weight between \textsf{HATT (unopt)} and \textsf{HATT} (up to 24 modes)}}
\resizebox{\linewidth}{!}{
\begin{tabular}{|l|l|l||l|l|l|}
\hline
\textbf{Case}   & \textbf{\textsf{HATT (unopt)}} & \textbf{\textsf{HATT}} & \textbf{Case} & \textbf{\textsf{HATT (unopt)}} & \textbf{\textsf{HATT}} \\ \hline
$H_2$ sto3g     & 32                             & 32                     & $3\times3$    & 404                            & 410                    \\ \hline
$LiH$ sto3g frz & 188                            & 188                    & $2\times5$    & 338                            & 330                    \\ \hline
$LiH$ sto3g     & 2880                           & 2850                   & $3\times4$    & 558                            & 524                    \\ \hline
$H_2O$ sto3g    & 5545                           & 5545                   & $3\times2F$   & 1266                           & 1290                   \\ \hline
$CH_4$ sto3g    & 37182                          & 37077                  & $3\times3F$   & 4976                           & 5153                   \\ \hline
$O_2$ sto3g     & 13082                          & 13370                  & $4\times2F$   & 3595                           & 3720                   \\ \hline
$2\times2$      & 82                             & 76                     & $4\times3F$   & 14330                          & 14267                  \\ \hline
$2\times3$      & 194                            & 187                    & $5\times2F$   & 7844                           & 7852                   \\ \hline
$2\times4$      & 261                            & 256                    & $6\times2F$   & 15005                          & 15047                  \\ \hline
\end{tabular}}
\vspace{-10pt}
\end{table}

\section{Related Works}

\textbf{Fermion-to-Qubit Mappings}: Previous Fermion-to-qubit mappings includes the Jordan-Wigner transformation~\cite{jordan1928uber}, Bravyi-Kitaev transformation~\cite{bravyi2002fermionic}, parity mapping~\cite{bravyi2017tapering} and ternary tree mapping~\cite{jiang2020optimal,miller2023bonsai}. The Bravyi-Kitaev transformation and ternary tree mapping generate the theoretical minimum Pauli weight per Majorana operator ($O(\log N)$). However, all of them fail to achieve optimal solutions with a specific Hamiltonian since the pattern of the Fermionic system is disregarded in their construction. Recent superfast encoding~\cite{setia2019superfast,riley2019analysis} captures structures of the Fermionic system to produce optimized mappings. However, it is only restricted to systems with local Fermionic modes and cyclic interaction patterns. Fermihedral~\cite{liu2024fermihedral} solves a given Hamiltonian's theoretical optimum Pauli weight mapping with an SAT problem. However, SAT is NP-complete, and Fermihedral fails to scale to cases larger than $\sim20$ qubits.
In contrast, \name{} combines the goal of optimizing Pauli weight and Hamiltonian information with the ternary tree construction process. It works on any Fermionic system and scales to larger cases with polynomial complexity $O(N^3)$.

\textbf{Quantum Simulation Compilers}: Some recent works successfully identify the computation pattern in quantum simulation to further optimize quantum circuits, including architectural-aware synthesis~\cite{arianne2020architecture, li2022paulihedral}, reordering Pauli strings for gate cancellation~\cite{gui2021term, hastings2015improving}, and simultaneous diagonalization~\cite{ewout2020circuit, alexander2020phase, cowtan2020generic}. These works happen after Fermion-to-qubit mappings and do not rely on a specific Fermion-to-qubit mapping.
\name~is compatible with these works.

\textbf{Quantum Compilers and Optimizations}: Modern quantum compilers, such as Qiskit~\cite{qiskit}, Cirq~\cite{cirq}, Braket~\cite{braket} and t$\ket{\text{ket}}$~\cite{sivarajah2021tket} apply multiple optimization passes to quantum circuits. These optimizations, including gate cancellations~\cite{nam2018automated} and rewriting~\cite{mathias2013white}, qubit routing~\cite{abtin2023qubit, murali2019noiseadaptive}, are applied after the Fermion-to-qubit mapping stage where the Hamiltonian simulation is converted to circuits.
Thus, they are compatible with \name{}.

\section{Conclusion}

In this work, we presented the Hamiltonian-Adaptive ternary tree (\name{}) compilation framework, a novel optimization for compiling Fermion-to-qubit mapping that leverages the high-level structure of the input Fermionic Hamiltonian. 
\name{} efficiently compiles ternary trees to optimize the mapping process, significantly reducing the Pauli weight and circuit complexity compared to existing methods. 
Our approach retains the \textit{vacuum state preservation} property and achieves a polynomial ($O(N^3)$) complexity, making it suitable for large-scale simulations that classical and exhaustive methods struggle to handle. 
Extensive evaluations demonstrated that \name{} not only achieves close-to-optimal mappings for small systems but also outperforms current constructive methods in larger-scale applications. 


\bibliographystyle{IEEEtranS}
\bibliography{refs}

\end{document}